\documentclass[aps,physrev,twocolumn,superscriptaddress, nofootinbib]{revtex4-2}

\usepackage{xcolor}
\usepackage{wrapfig}
\usepackage{hyperref}
\usepackage{subcaption}
\usepackage{amsmath}
\usepackage{graphicx}

\usepackage{tabularx} 
\newcolumntype{Y}{>{\centering\arraybackslash}X}
\usepackage{dcolumn}
\usepackage{float} 
\usepackage{bm}
\usepackage[export]{adjustbox}
\usepackage{multirow} 
\usepackage{siunitx}
\usepackage[utf8]{inputenc}
\DeclareUnicodeCharacter{00B3}{\(^3\)} 
\DeclareUnicodeCharacter{202F}{\,}

\begin{document}
\setlength\parindent{30pt}
\preprint{Preprint}

\title{A Machine Learning Assist for Core-Collapse Supernovae Gravitational Wave Detection}

\author{Andy H.Y. Chen}
    \affiliation{%
        Institute of Physics, National Yang-Ming Chiao Tung University, Hsinchu, Taiwan %
    }%
    \email{%
    contact author: andy.c.80297@gmail.com
    }%
\author{Chia-Jui Chou}
    \affiliation{%
        School of Physical Science and Technology, ShanghaiTech University, China %
    }%
\author{Kuo-Chuan Pan}
    \affiliation{%
        Institute of Astronomy, National Tsing Hua University, Hsinchu, Taiwan %
    }%
\author{Yi Yang}
    \affiliation{%
        School of Physical Science and Technology, Center for Fundamental Physics, ShanghaiTech University, P. R. China %
    }%
\author{Shih-Chieh Hsu}
    \affiliation{%
        Department of Physics, University of Washington, Seattle, USA %
    }%
\author{Albert K.H. Kong}
    \affiliation{%
        Institute of Astronomy, National Tsing Hua University, Hsinchu, Taiwan %
    }%
    \affiliation{%
        Institute of Space Engineering, National Tsing Hua University, Hsinchu, Taiwan %
    }%
\author{En-Tzu Lin}
    \affiliation{%
        Institute of Astronomy, National Tsing Hua University, Hsinchu, Taiwan %
    }%
\author{Lupin C.-C. Lin}
    \affiliation{%
        Department of Physics, National Cheng Kung University, Tainan 701401, Taiwan %
    }%
\author{Yu-Chiung Lin}
    \affiliation{%
        Institute of Astronomy, National Tsing Hua University, Hsinchu, Taiwan %
    }%
\author{Li-Ting Ma}
    \affiliation{%
        Institute of Astronomy, National Tsing Hua University, Hsinchu, Taiwan %
    }%
\author{Surojit Saha}
    \affiliation{%
        Institute of Astronomy, National Tsing Hua University, Hsinchu, Taiwan %
    }%
\author{Sut-Ieng Tam}
    \affiliation{%
        Institute of Physics, National Yang-Ming Chiao Tung University, Hsinchu, Taiwan %
    }%
\author{Shu-Wei Yeh}
    \affiliation{%
        Department of Physics, National Tsing Hua University, Hsinchu, Taiwan %
    }%

\date{\today}%
\begin{abstract}
Gravitational-wave astronomy has achieved remarkable success in detecting compact binary coalescences. 
However, detecting core-collapse supernovae (CCSNe) remains challenging because their gravitational-wave strains are expected to be weak at ground-based detectors, and the waveform information can be contaminated by transient noise artifacts (glitches) present in the detector. In this work, we present \texttt{CCSNet}, a machine learning–based binary classification framework designed to identify CCSN signals in observational strain data by distinguishing them from glitches. We trained ML models with different CCSN injection configurations using 31 three-dimensional self-consistent core-collapse supernova gravitational simulations. The variation in injection configurations includes the selection of the simulation template group, signal durations, and the number of detectors. After training, we test the CCSN sensitivity under different types of background regarding the existence of glitch signals. We assess the sensitivity of the model when 50\% of injected signals were recovered with a 95\% glitch rejection rate. Compared to the background without glitch signals, our model preserves approximately 92\% of the signal-to-noise ratio (SNR) sensitivity when glitch signals overlap with CCSN signals. Moreover, the glitch mitigation behavior for CCSN detection can be extended to single-detector data, where approximately 75\% of the SNR sensitivity is preserved compared to the dual-detector setup. These results demonstrate that \texttt{CCSNet} can be used to search for CCSN signals when the detector is not in an ideal situation for traditional detection methods. 
\end{abstract}
\maketitle

\pagebreak
\section{Introduction} 

Core-collapse supernovae (CCSNe) are violent explosions of massive, dying stars that emit gravitational waves (GW), electromagnetic radiation, and potentially detectable neutrino counterparts \cite{PhysRevLett.58.1494, PhysRevLett.58.1490, 2306.08024}. The GW from CCSNe carries crucial information about the dynamics of their central engines, such as the collapse of stellar cores and extreme nuclear matter behavior. These GW signals are expected to reach observatories before their electromagnetic counterparts \cite{1908.03584}, providing an early-warning signal or multi-messenger observations, albeit with very weak GW strains. The designed sensitivity of Advanced LIGO \cite{1411.4547}, Advanced Virgo \cite{1408.3978}, and KAGRA \cite{1811.08079} (LVK) detectors allows for the detection of CCSNe events within the Milky Way or, in extreme cases, up to tens of Mpc \cite{1605.01785, 1908.03584, 2305.16146}. However, all GW events detected to date \cite{1811.12907, 2010.14527, 2108.01045, 2111.03606} belong to the category of compact binary coalescence (CBC) systems. 

The GW detection pipeline can be divided into template-based~\cite{1908.11170, 9908043} and non-template-based search based on the reliance on waveform models. Template-based searches rely on a large amount of accurately modeled GW waveform templates to identify similar patterns in the detector through matched-filtering. This approach is well-suited for CBC waveforms, but not for CCSN, where their waveform cannot be well-modeled \cite{1206.2503, 2306.08024}. On the other hand, non-template-based methods, also known as burst search, detect the GW with minimal or no waveform assumptions about the waveform morphology, allowing for the identification of unmodeled or unexpected sources beyond CBC events. 

Although the GW detectors are calibrated to produce well-characterized data, environmental disturbances and quantum noise can introduce non-Gaussian noise transients, known as glitches. These glitches can hinder the detection pipelines and lead to false alarms. To mitigate such false triggers, pipelines must determine whether the candidate events belong to a noise artifact and veto those signals, preventing the pipeline from producing false alarms. To reduce the false alarm \cite{1604.04324, 1508.02357, 1512.02864, 2112.03597}, a variety of glitch-veto and noise-characterization algorithms have been applied to the detection pipeline. 
However, these methods only apply to CBC waveforms where a large amount of accurately modeled waveforms are available and are not directly optimized for other types of astrophysical signals. 

Machine-learning approaches have been increasingly applied to gravitational-wave data analysis~\cite{2005.03745, 2412.15046}, offering data-driven alternatives to traditional analysis methods. In CBC searches, \texttt{Aframe}~\cite{2403.18661} employs a ResNet \cite{1512.03385} model for CBC detection, followed by \texttt{AMPLFI}\cite{2407.19048}, which performs low-latency sky-localization inference. For burst searches, \texttt{MLy}\cite{2009.14611} and \texttt{GWAK}\cite{2309.11537, 2412.19883} utilized non-supervised ML models to detect GW signals without relying on predefined waveform templates. Besides form CBC and general burst searches, several supervised CCSN detection~\cite{1812.05363, 1912.13517, 2011.13733, 2411.12524} ML models demonstrated promising sensitivity on CCSN injection in their test dataset. Supervised ML models can also be trained to classify CCSN~\cite{2001.00279, 2009.07367, 2209.10089, 2301.09387, 2304.11498, 2310.09551, 105.084054} based on the properties of the progenitor star. These supervised models can offer an alternative approach for CCSN detection when the model is robust. Moreover, ML methods are also used to identify glitch signals in gravitational-wave data. \texttt{GWSkyNet}~\cite{2010.11829} discriminates glitch signals for CBC search during online runs when multi-detector data is provided, and \texttt{GravitySpy}~\cite{1611.04596, 2308.15530} utilizes convolutional neural network (CNN) classifiers to label the glitch signals based on their morphologies during offline analysis. The supervised ML classifier models for CCSN and glitches ~\cite{2002.04591, 2111.07219} can be used to reduce the distibution of noise triggers that overlap with CCSN triggers using one or more detector data. Collectively, these developments demonstrate the broad utility of ML across GW detection and data characterization, establishing it as a versatile framework for identifying, classifying, and distinguishing both astrophysical and non-astrophysical transients. 

Given that the waveforms of CCSNe and detector glitches are difficult to model accurately, searches for CCSN gravitational-wave signals rely on non-template-based methods. These searches require coincident strain data from at least two detectors, which reduces the effective observing duty cycle. Furthermore, the absence of waveform templates makes the identification of candidate events as a glitch or CCSN a challenging task. To address these limitations, we develop \texttt{CCSNet}\footnote{\url{https://github.com/AndyC80297/CCSNet}}, a supervised machine-learning–based transient classification framework designed to distinguish glitch artifacts from a broad range of CCSN signals. The framework demonstrates the ability to identify CCSN waveforms not included in the training dataset, highlighting its generalization capability. In addition, we show that \texttt{CCSNet} can recover CCSN signals from glitch backgrounds, generate triggers based solely on the early evolution of CCSN waveforms, and operate in single-detector configurations. Using O3b science-run data, \texttt{CCSNet} achieves a 50\% detection efficiency at a distance of 3.89 kpc with a 5\% false-positive rate, demonstrating its potential to enhance both CCSN searches and glitch-mitigation studies. 

In the following sections, we describe the process of data preparation for training, validation, and testing, together with the properties of the CCSN waveform in Section \ref{sec_data}. Section \ref{sec_training} provides an overview of our model architecture and details the training process of the deep neural networks. Section \ref{sec_result} presents the results of various trained models. In Section \ref{sec_disscussion}, we analyze and compare the test performance of the different models. Finally, in Section \ref{sec_conclusion}, we summarize our conclusions and outline future directions for this work. 

\section{Data \label{sec_data}}

We prepare noise and injection data to train our supervised model. Each training sample consists of a strain segment paired with a label indicating whether it contains pure noise or an injected CCSN signal. Using this dataset structure, we train multiple models with different CCSN injection configurations and compare their performance in identifying CCSN waveforms that are not included in the training or validation sets. 

\begin{figure*}[tb]
  \centering 
  \includegraphics[width=1\linewidth]{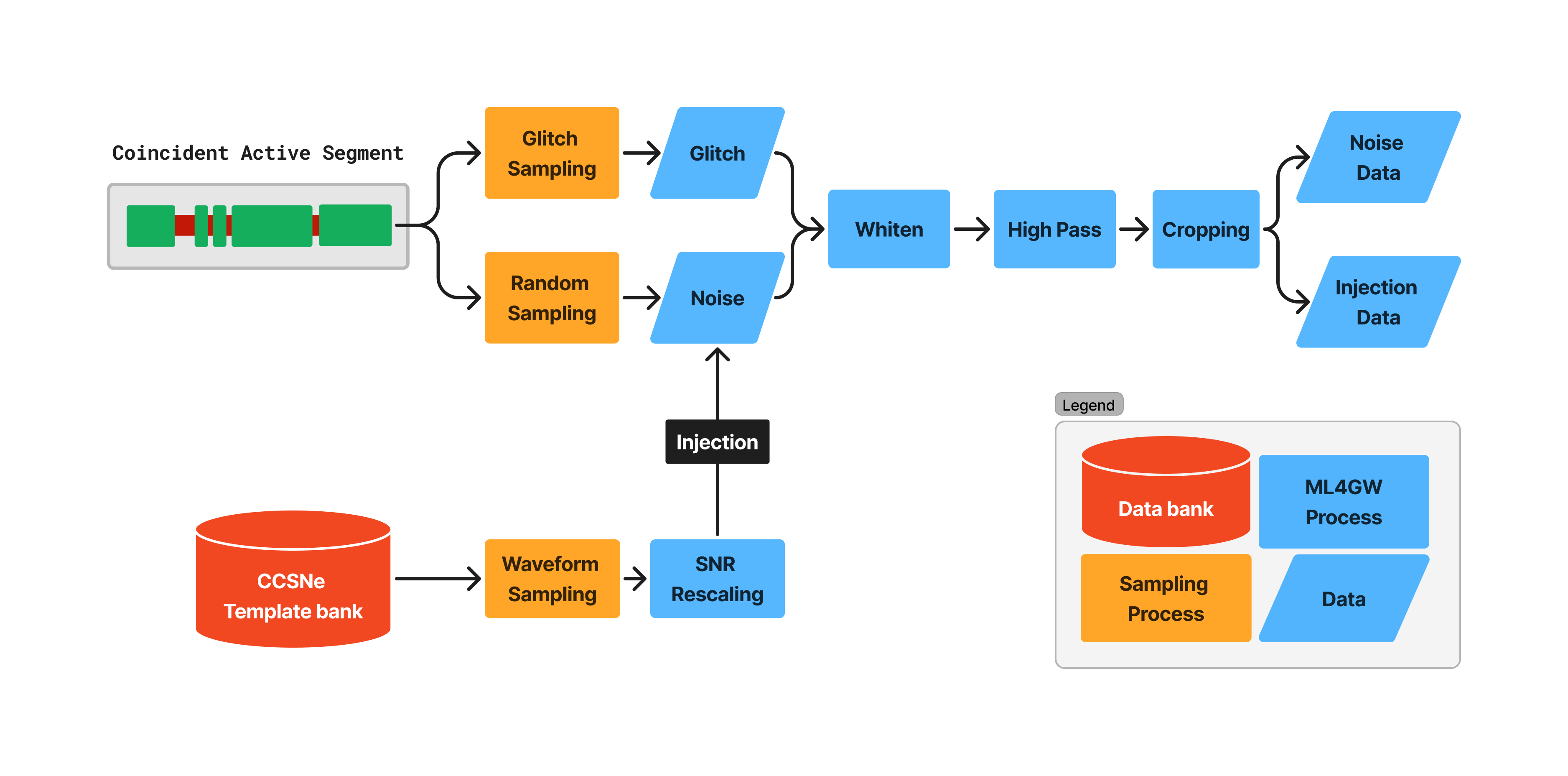}
  \caption{The diagram illustrates the workflow for data generation. In each iteration, \texttt{CCSNet} samples new backgrounds and CCSN waveforms for training. Dynamic processes, including distance rescaling and waveform sampling, are highlighted within the yellow square block. This dynamic approach ensures a more robust training strategy compared to relying on a fixed dataset. 
  }
  \label{fig: train_data_workflow}
\end{figure*}

We follow the workflow in Figure \ref{fig: train_data_workflow} to prepare our dataset. The training dataset contains two channels of one-dimensional time series data, specifically strain data from LIGO Hanford (H1) and Livingston (L1). We whiten both noise and injection data using the power spectral density (PSD) estimated from the full sampled segment. The PSD calculation follows the Welch method \cite{welch2003use}, employing a two-second frequency window with one-second overlap, and a Hann window function to mitigate spectral leakage. After whitening, a 32 Hz high-pass filter is applied to remove low-frequency noise, and one second of data is removed from both ends to reduce edge effects from the discrete Fourier transform. The following subsections describe the preparation of strain data and CCSN templates, and then outline the process for producing the training, validation, and test datasets. 

\subsection{O3b Detector Strain Noise} 

We conduct our analysis using H1 and L1 strain data from the O3b run, specifically the DCS-CALIB\_STRAIN\_CLEAN\_C01\_AR dataset that meets the DCS-ANALYSIS\_READY\_C01 data quality flag. This choice ensures access to scientific-grade data suitable for our search studies. To access a sufficient amount of glitch data for both model training and validation, we selected an O3b segment that is longer than 30,000 seconds for data sampling. For testing, we sampled segments longer than 15,000 seconds from the subsequent four weeks. In this study, the segment between 2020-01-10 01:10:36 and 2020-01-10 10:16:44 UTC, providing 32,768 seconds of data, is used for training and validation, while 31 additional segments longer than 15,000 seconds from the following four weeks are used for testing. This selection enabled us to analyze our model with a substantial amount of data. The entire period of the strain data for training, validation, and testing spans from 2020-01-10 01:10:36 to 2020-02-07 09:03:38 UTC. Testing on data segments distinct from those used in training enables us to assess the performance of our model under varying detector conditions. The strain is downsampled from $16384$ Hz to $4096$ Hz for computational and training efficiency. 

To identify the timestamps of glitch signals, we pass the strain data through the Omicron software package \cite{2007.11374}, using the parameters listed in Table \ref{omicron_table}. The Omicron software applies a multi-Q transform to the time-series data and computes the transform coefficient $X$ Eq. \ref{coefficient}, which measures the average signal amplitude and phase within a localized time–frequency region using a Gaussian window function $w$. From this coefficient, the q-transform signal-to-noise ratio (SNR) $\hat{\rho}(\tau,\phi, $Q$)$ Eq. \ref{qt_snr} is constructed. Timestamps with q-transform SNR greater than 5 are labeled as glitch signals for our analysis, where the $\tau$, $\phi$, and $Q$ represent time, phase, and the constant-Q transform value \cite{1.400476}, respectively.

\begin{equation}
    \label{coefficient}
        X(\tau, \phi, Q) = \int_{a}^{b} x(t) w(t-\tau) e^{-2i \pi \phi t} dt\\
\end{equation}

\begin{equation}
    \label{qt_snr} 
        \hat{\rho}(\tau,\phi,Q) =\\
         \begin{cases}
            \sqrt{\left|X^{wh}(\tau, \phi, Q)\right|^2-2}, & \mathrm{if}\ \left|X^{wh}\right|^2 \ge 2\\
            0, & \mathrm{otherwise}.
        \end{cases}
\end{equation}

\begin{table}[tb]
    \begin{tabularx}{\linewidth}{ c  Y  Y }   \hline\hline
        Parameters & value & Units\\              \hline 
        Q-value range  & (3.1328,  108) & - \\    
        Frequency range  & (32, 2048) & Hz\\      
        Chunk duration  & 128 & second \\         
        Segment duration  & 64 & second\\         
        Overlap duration  & 4 & second \\         \hline\hline
        \end{tabularx}
\caption{List of criteria to pick up short transients via Omicron.}
\label{omicron_table}
\end{table}

\subsection{Core-collapse Supernova Waveforms}

Thanks to the recent advancements in supercomputer facilities, realistic CCSN simulations incorporating three-dimensional (3D) radiation hydrodynamics, self-consistent neutrino transport, and general relativistic effects have become possible. In this study, we consider 31 gravitational waveforms from state-of-the-art 3D CCSN simulations that have been previously studied in \cite{2104.06462} and \cite{2305.16146}. These waveforms are taken from \cite{1607.05199, 1810.07638, 1807.07579, 2002.10115, 1812.07703, 2010.02453} and are listed in Table~\ref{ccsn_table}. Note that our collection includes a wide variety of CCSN simulations, which differ by several key parameters: simulation duration (either from core collapse or several milliseconds after core bounce), progenitor mass, nuclear equation of state (EoS), and rotation.

Table \ref{ccsn_table} lists the parameters of these waveforms, including progenitor~\cite{0702176, 1510.04643, 10.3847/1538-4357/aac2da} masses, spins, and EoS, which includes both typical and extreme CCSN waveforms.  The extreme CCSN waveforms in our template bank include Pan 2021 SR, Pan 2021 FR, and Powell 2020 m39, which feature fast rotation or strong magnetohydrodynamic effects at higher masses. We converted these simulation results into the second-derivative quadrupole moment (SDQM) representation and resampled them to 4096 Hz to match the sampling rate of the background data. 

\begin{center}
    \centering
    \begin{table*}[tb]
        \begin{tabularx}{\linewidth}{c  Y  Y  Y  Y  Y  Y }
            \hline\hline
            Paper & Name & Mass $(M_{\odot})$ & Rotation & EoS & Start time (ms) & Duration (ms) \\
            \hline
            Andresen 2016 \cite{1607.05199}  & s11 & 11.2 & False & LS220 \cite{LS220-paper} & 11 & 335 \\ 
            & s20 & 20 & False & LS220 & 7 & 415 \\ 
            & s20s & 20 & False & LS220 & 10 & 519 \\ 
            & s27 & 26 & False & LS220 & 10 & 564 \\ 
            \hline
            Andresen 2019 \cite{1810.07638} & s15nr & 15 & False & LS220 & 10 & 319 \\ 
               & s15sr & 15 & True & LS220 & 10 & 358 \\ 
               & s15fr & 15 & True & LS220 & 11 & 450 \\ 
            \hline
            O'Connor 2018 \cite{1807.07579} & mesa20$^{*}$ & 20 & False & SFHo & 16 & 485 \\ 
              & mesa20\_pert$^{*}$ & 20 & False & SFHo & 16 & 512 \\ 
              & mesa20\_LR & 20 & False & SFHo & 16 & 621  \\ 
              & mesa20\_v\_LR$^{*}$ & 20 & False & SFHo & 16 & 468 \\ 
              & mesa20\_pert\_LR & 20 & False & SFHo & 16 & 503 \\ 
            \hline
            Pan 2021 \cite{2010.02453} & NR$^{*}$ & 40 & False & LS220 & -471 & 1247 \\ 
              & SR$^{*}$ & 40 & True & LS220 &  -474 & 1410 \\ 
              & FR$^{*}$ & 40 & True & LS220 & -485 & 943 \\ 
            \hline
            Powell 2020 \cite{2002.10115} & s$18^{*}$ & 18 & False & LS$220^{*}$  & -318 & 882 \\ 
              & y$20^{*}$ & 20 & False & LS$220^{*}$  & 1 & 1205 \\ 
              & m$39^{*}$ & 22 & True & LS$220^{*}$ & 2 & 1004 \\ 
            \hline
            Radice 2019 \cite{1812.07703} & s09 & 9 & False & SFHo & 10 & 1032 \\ 
              & s10 & 10 & False & SFHo & 10 & 757 \\ 
              & s11 & 11 & False & SFHo & 10 & 558 \\ 
              & s12 & 12 & False & SFHo & 10 & 893 \\ 
              & s13 & 13 & False & SFHo & 10 & 761 \\ 
              & s14 & 14 & False & SFHo & 10 & 984 \\ 
              & s15 & 15 & False & SFHo & 10 & 659 \\ 
              & s16 & 16 & False & SFHo & 10 & 606 \\ 
              & s17 & 17 & False & SFHo & 10 & 639 \\ 
              & s18 & 18 & False & SFHo & 10 & 608 \\ 

              & s19 & 19 & False & SFHo & 10 & 861 \\ 
              & s20 & 20 & False & SFHo & 10 & 619 \\ 
              & s60 & 60 & False & SFHo & 
              10 & 388 \\ 
            \hline\hline
        \end{tabularx}
    \caption{The table lists the progenitor mass, rotation configuration, equation of state (EoS), start time, and simulation end time of the CCSN templates used in this study. Templates marked with an asterisk (*) are selected for the sub-template bank. We label $t=0$ as the core-bounce time, and the duration corresponds to the post-bounce portion of the signal. Among these, simulations that employ the LS$220^{*}$ EoS adopt the low-density treatment described in \cite{RamppJanka2002}.} 
    \label{ccsn_table}
    \end{table*}
\end{center}

\subsection{Data sampling}

To efficiently generate training data with different injection configurations, we produce new datasets in each training iteration. This on-the-fly data generation process enables us to reduce disk usage and train our model with a larger amount of data. This type of data, also known as dynamic data, is typically generated in real time or updated frequently. In contrast, static datasets remain fixed once collected, representing a snapshot of a system at a specific point in time. Training models on dynamic data forces them to handle evolving patterns, reducing the risk of overfitting that often arises with static datasets. Consequently, this approach enhances the ability of the model to generalize in non-stationary environments, leading to more accurate predictions and decisions. 

However, generating large volumes of dynamic data in real time can be computationally expensive. To overcome this challenge, we employ \texttt{ml4gw}\footnote{\url{https://github.com/ml4gw/ml4gw}} \cite{ml4gw}, a PyTorch-based library for time series data augmentation in machine learning applications. \texttt{ml4gw} leverages the parallel processing power of GPUs to manage data generation and augmentation tasks efficiently. We follow the workflow illustrated in Figure \ref{fig: train_data_workflow} to sample and process the data before sending it into our model. The following sections provide a detailed explanation of the sampling process. 

\subsubsection{Noise Data}

We randomly sample three seconds of strain data from each detector independently to generate the noise dataset. The sampled stain data includes both glitch and non-glitch signals from the H1 and L1 detectors. By combining the strain data from the two detectors, we can categorize them into four types of noise. Below, we list the portion of each type of background used for training and validation. 

\begin{itemize}
    \item Stationary: Noise data sampled from H1 and L1 without glitch signals in either detector. Pure stationary-noise segments without injections are excluded from the training dataset, so that the model focuses on discriminating astrophysical signals from non-stationary transient noise artifacts. However, stationary noise is still used as the background for signal injections during training. We additionally include pure stationary-noise samples in the test set to evaluate the model performance on noise below the Omicron triggering threshold. 
    \item H1-Glitch: Noise data containing glitches in H1, with no glitches in L1. This constitutes 37.5\% of the training noise dataset and 25\% of the validation noise dataset. 
    \item L1-Glitch: Noise data containing glitches in L1, with no glitches in H1. This also constitutes 37.5\% of the training noise dataset and 25\% of the validation noise dataset. 
    \item Combined: Noise combination of glitch signals from H1 and L1 detectors, taking up 25\% of both the training and validation noise datasets. 
\end{itemize}

In each training iteration, we sample 102,400 noise data, with half allocated to glitch data and the other half used for injections. During validation, we sample 1,024 noise data points to evaluate the performance of our model on different background types. This validation noise data is also used for CCSN injections during the validation phase. 

\subsubsection{Injection Data}
\begin{figure*}[tb]
  \centering
  \includegraphics[width=1\linewidth]{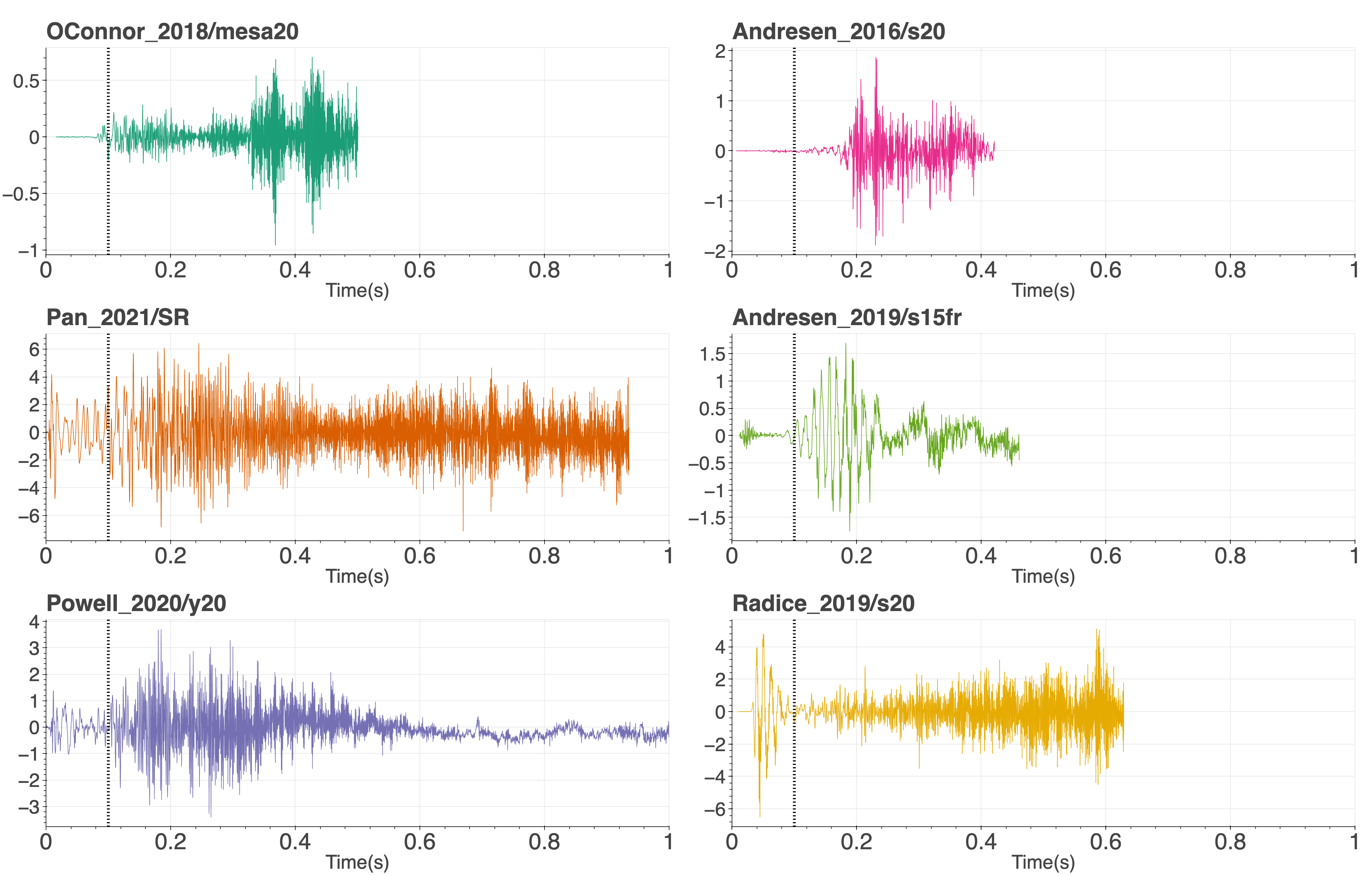}
  \caption{CCSN $h_{+}$ polarization at the equator orientation with distance at $10$~kpc. The amplitude in the plot is rescaled by a factor of $10^{22}$ for an easier view. The dashed line indicates the 100 milliseconds after core-bounce time, which are the signals used for models that train only on early waveforms.}
  \label{fig: CCSN_waveforms}
\end{figure*}

We generate the injection data by adding simulated CCSN waveforms into the sampled noise. We impose four injection configurations based on the combinations of the two injection restrictions: template selection and waveform truncation. In the first restriction, we select nine CCSN templates from our 31 templates to form our sub-template bank. We select waveform families that have simulations starting before the core-bounce time to our sub-template bank, which includes templates from Powell \cite{2002.10115} and Pan \cite{2010.02453}. Additionally, we include waveforms with higher resolution in O'Connor \cite{1807.07579} to enrich the EoS of our sub template bank for training, which are mesa20, mesa20\_pert, and mesa20\_v\_LR. Note that although all the Eos of the waveforms in O'Connor and Radice belong to SFHo\cite{SFHo-paper}, we decide to train on a smaller set of waveforms and test our robustness with a larger set of CCSN. So we flavor O'Connor over Radice \cite{1812.07703} for our sub template bank selection. The selected waveforms are labeled with a star sign in the Name row in Table \ref{ccsn_table}, which spans across a range of parameters such as mass, spin, and EoS. The second restriction is waveform truncation, which we apply to the waveforms at 100 milliseconds after core-bounce time, as in Figure \ref{fig: CCSN_waveforms}. We call the waveform that has been truncated an early waveform; otherwise, they are called full waveforms. This truncation allows us to focus on the GW signal around the core-bounce phase. By combining these two injection constraints, we form four different injection configurations for model training. We named the model trained under such configurations as the descriptions in Table \ref{model_table}. 

\begin{enumerate}
    \item Model\_FF: The model that was trained on full simulation injections of all 31 waveforms. Since this model is trained on the full CCSN information provided in this study, we use this model as the baseline model for performance comparison. 
    \item Model\_SF: The model that was trained on full simulation injections of the sub-waveform. 
    \item Model\_FE: The model that was trained on early simulation injections of all 31 waveforms. 
    \item Model\_SE: The model that was trained on early simulation injections of the sub-waveform. 
\end{enumerate}

\begin{table}[tb]
    \begin{tabular}{ c c c}
    \hline\hline
                   & Full template bank & Sub template bank \\ \hline
    Full waveform  & Model\_FF    & Model\_SF   \\ 
    Early waveform & Model\_FE    & Model\_SF   \\ \hline\hline
    \end{tabular}
    \caption{Summary of injection methods used for training. The table categorizes models based on the injection constraint applied during training.}
    \label{model_table}
\end{table}

The CCSN template is injected into noise randomly sampled from the strain data of the L1 and H1 detectors. In each iteration, we randomly sample 51,200 CCSN templates for training. For each template, we sample different orientations, phases, sky localizations, and time shifts according to the parameters outlined in Table \ref{ccpa_table}. Note that the time shifts refer to the core-bounce offset from the center of the input kernel in both detectors. This allows us to preserve the physical time delay of the two detectors due to the sky location and add data augmentation for model training. To enhance the diversity of our injection data, we rescale the CCSN distance to match a network SNR $\rho_{net}$ (Eq. \ref{net_snr}) to a distribution that follows a power law with an index of 3 within the range of 12 to 100. It is important to note that the q-transform SNR differs from the network SNR.

\begin{equation} 
    \label{net_snr}
        \begin{split}
        \rho^{2}_{det} =
            4 \int_{f_{\text{min}}}^{f_{\text{max}}}
            \frac{\tilde{h_{}}(f)\tilde{h_{}}^*(f)}
            {S_n^{}(f)}df \\
        \rho_{net} = \sqrt{\sum^{n}_{i=0}(\rho^{2}_{det})_{i}}
        \end{split}
\end{equation}

The network SNR is computed as the square root of the sum of the matched-filtered SNR $\rho^{2}_{det}$ from each detector. To estimate the matched-filtered SNR, we integrate the injected signal from $f{\text{min}}$ = 32 Hz to $f_{\text{max}}$, which is bounded by the Nyquist frequency. Here, $\tilde{h}(f)$ represents the Fourier transform of the time-domain CCSN gravitational wave signal, while $S_n(f)$ denotes the amplitude spectral density of the background, derived from the square root of the PSD. 

During validation, we sample 1,024 waveforms for each CCSN template used in training, with parameters drawn from the priors listed in Table~\ref{ccpa_table} and $\rho_{\text{net}}$ values ranging from 8 to 12. The $\rho_{\text{net}}$ distribution of the training and validation datasets is intentionally different: training samples have $\rho_{\text{net}}$ between 12 and 100, while validation samples cover the lower range. This design allows the model to learn CCSN features from high-SNR injections without being dominated by noise contamination, while the lower-SNR validation data are used to assess the classification performance when the injection signals are deeply buried. After sampling, the validation data are processed using the same procedure applied to the training data, as illustrated in Figure~\ref{fig: train_data_workflow}, before being fed to the model. 

\begin{center}
    \begin{table}[tb]
        \begin{tabularx}{\linewidth}{cYYY}
        \hline\hline
        Parameter                                                                   & Prior   & Limits              & Units  \\ \hline
        \begin{tabular}[c]{@{}c@{}}Orientation \\ Declination\end{tabular}          & Cosine  & ($-\pi$/2, $\pi$/2) & rad.   \\ 
        \begin{tabular}[c]{@{}c@{}}Orientation \\ Right ascension\end{tabular}      & Uniform & (0, 2$\pi$)         & rad.   \\ 
        \begin{tabular}[c]{@{}c@{}}Sky localization \\ Declination\end{tabular}     & Cosine  & ($-\pi$/2, $\pi$/2) & rad.   \\ 
        \begin{tabular}[c]{@{}c@{}}Sky localization \\ Right ascension\end{tabular} & Uniform & (0, 2$\pi$)         & rad.   \\ 
        Polarization Phase                                                          & Unifrom & (0, $\pi$)          & rad.   \\ 
        Time shift                                                                  & Uniform & (-0.2, 0.1)         & second \\ \hline\hline
        \end{tabularx}
    \caption{The parameters used for waveform projections in each iteration. Due to the asymmetric structure of 3D simulations, waveforms exhibit different behaviors depending on the explosion orientation.}
    \label{ccpa_table}
    \end{table}
\end{center}

\subsection{Test dataset}
We prepare all four types of noise data from the test segments, including injections of different types of CCSN, to evaluate our model. The test noise includes 20,000 background samples for each noise type for each test segment. To create a more comprehensive test set, we generate injections with varying intensities based on network SNR values or distances. Since CCSN waveforms depend strongly on progenitor properties, different templates can produce substantially different signal amplitudes and network SNRs even at the same distance. Injecting signals at a fixed network SNR provides a controlled way to evaluate the model performance with a consistent signal amplitude relative to the background noise, independent of the intrinsic explosion strength of individual CCSN templates. This allows us to assess the average sensitivity of the model across different CCSN waveform morphologies. In contrast, injections at fixed distances preserve the intrinsic amplitude variations among CCSN models and therefore provide a more astrophysically motivated evaluation. Comparing detectability across templates at different distances helps identify which CCSN signals are more detectable within our framework and provides insight into the generalizability and astrophysical relevance of the model. 

To assess the performance of our models, we inject CCSN signals with eleven network SNR intensities ranging from 4 to 40. The lower bound allows us to examine our CCSN sensitivity at nearly undetectable injection intensity, while the upper bound evaluates the performance of the models on exceptionally strong injections. We also sample twenty logarithmically spaced injection distances between the minimum and maximum luminosity distances. The maximum injection distance is set to twice the average luminosity distance, where the CCSNe have a network SNR equal to 4, while the minimum distance is 100 times smaller than the maximum. Due to variations in the astrophysical properties of progenitor stars, the minimum and maximum distances differ across CCSN templates. Since the CCSN templates used in this study are derived from 3D GRHD simulations, explosion orientation significantly affects waveform amplitude. Therefore, a broader range of injection distances ensures coverage of the region of maximum detection efficiency, accounting for variations in detectability due to orientation-dependent amplitude differences. 

For each CCSN template, we generate 20,000 pairs of polarizations, with priors and boundaries defined in Table \ref{ccpa_table}. The noise and injection data are normalized using the PSD estimated from the sampled segments. The test PSD is calculated using the same method as the training PSD, and the test data undergoes identical preprocessing steps, including normalization, high-pass filtering, and cropping, before being sent into the ML model. In total, we produce 2,698,240,000 test samples, consisting of four types of noise data from 31 segments, with CCSN injections from 31 simulations across different SNRs and distances.

\section{CCSNet and training procedure \label{sec_training}}

\subsection{CCSNet framework}

The \texttt{CCSNet} framework is designed to streamline the data preparation and model training for CCSN search. It integrates three key components: strain data fetching, automated Omicron scan for glitch identification, and on-the-fly data augmentation during model training. The framework automatically retrieves strain data from the detector with the given time segment and applies the Omicron scan to identify transient noise features. During training, the framework dynamically constructs training batches that inject simulated CCSN into real detector backgrounds. This automation ensures consistent data processing and efficient utilization of computing resources across large datasets. 

Beyond architectural or optimization considerations, \texttt{CCSNet} provides a framework to evaluate the influence of training data configurations on model performance. By systematically varying injection strategies, waveform selections, and detector combinations, the framework enables controlled comparisons of model robustness and generalization. This design allows for the rapid datasets generation, facilitating scalable experimentation and reproducible benchmarking in CCSN detection studies. Within this framework, multiple models are trained following the procedure described in the subsequent section. 

\subsection{Training procedure}
The NN model used in our analysis is based on NN architectures that have demonstrated success in processing strain data and extracting CCSN features. The convolutional neural network (CNN) classifier in this study~\cite{2001.00279} achieved promising results in detecting CCSN injections on simulated LIGO background noise. To enhance model robustness against false alarms caused by glitch signals in real scientific runs, we increased the model complexity by replacing standard CNN layers with dilated CNN layers, as proposed in \cite{1904.08693, 1609.03499}, and made adjustments to match our input data. The study in \cite{1904.08693}, which focused on gravitational wave reconstruction using LIGO O2 data, demonstrated that dilation features effectively mitigate the impact of glitch signals in scientific runs. 

Our deep learning model consists of twelve dilated CNN layers, followed by a CNN reshaping layer, two fully connected layers, and a single-node output layer. We use the ReLU activation function to facilitate efficient training between layers. To improve training stability, we incorporate group normalization \cite{1803.08494}, batch normalization \cite{1502.03167}, and Kaiming initialization \cite{1502.01852}. During training, we optimize the NN parameters to minimize binary cross-entropy loss. We select AdamW \cite{1711.05101} as our optimizer instead of the standard Adam \cite{1412.6980} to apply effective L2 regularization, which helps prevent overfitting. To accelerate convergence, we initialize the learning rate at 0.01 and adjust it dynamically using a one-cycle learning rate scheduler with cosine annealing \cite{1708.07120}. 

We train the model on a single NVIDIA A100 80GB GPU with a batch size of 512, balancing computational efficiency and convergence speed. The model undergoes 200 training iterations, each comprising 200 batches of data. In total, the model processes 20,480,000 training samples over approximately four hours. After training, if no signs of overfitting are observed, we use the final-iteration model for the test analysis; otherwise, we select the checkpoint model with the best validation performance after the 100th iteration. 

\begin{figure}[tb]
    \includegraphics[width=0.95\linewidth]{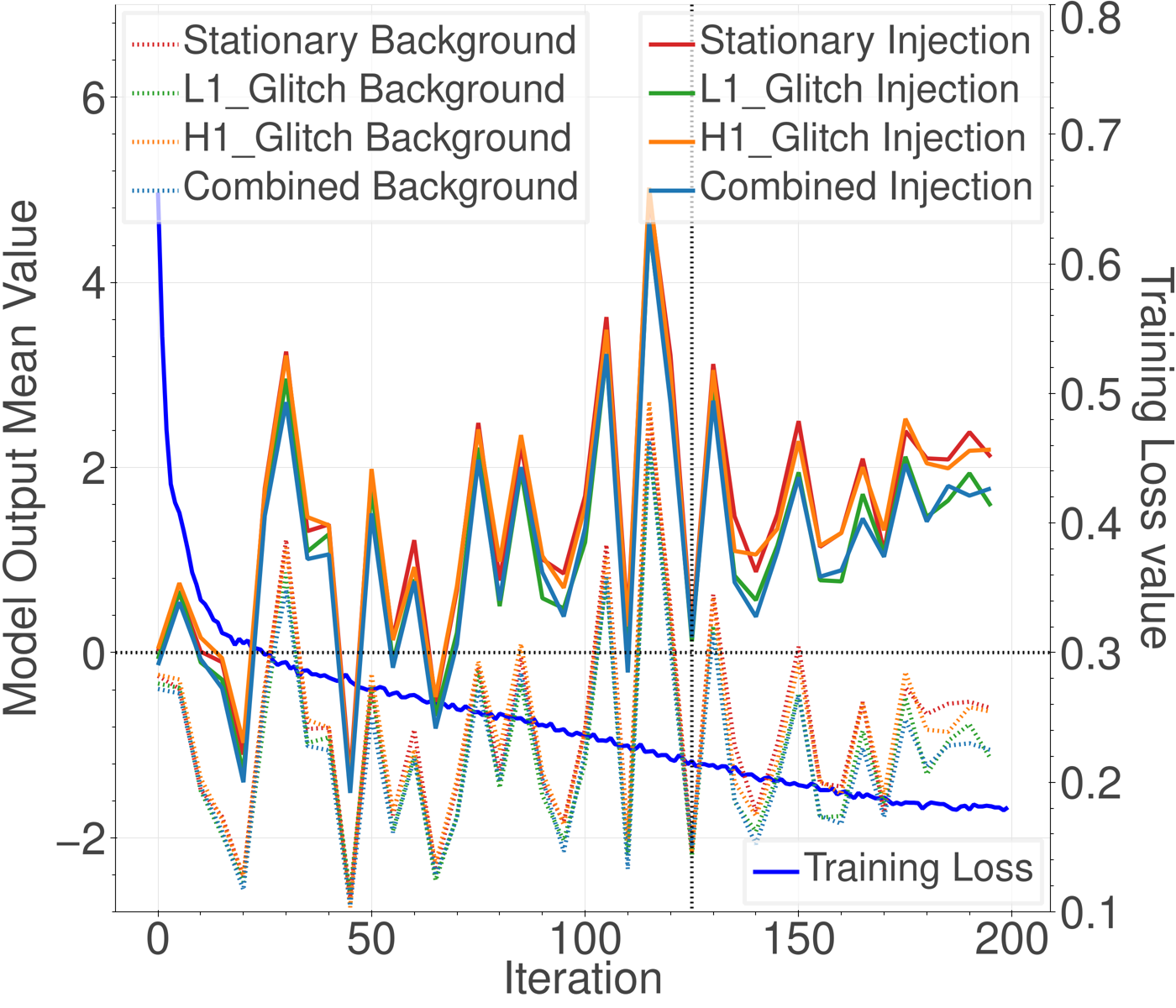}
    \caption{
        This is the training and validation performance of the Model\_FF. 
        This figure shows the model output mean value across different background and injection types, overlaid with the training loss (dark blue line). The vertical dashed line marks iteration 125, which was selected as the final model according to the criteria described in our validation methodology.
    }
    \label{validation}
\end{figure}
\subsection{Validation methodology}
Overfitting is typically evaluated by examining discrepancies between training and validation losses. However, the loss function provides only a single scalar measure of classification performance between noise and CCSN injections, and therefore cannot fully interpret the robustness of the model. Moreover, contributions from the environmental noise source may change over time, which introduces new types of glitches. So we can not assess the robustness of our model based solely on the loss.  

Figure~\ref{validation} shows the training loss and validation performance of Model\_FF, illustrating the general convergence behavior of our models. To assess robustness, we evaluate the model using the noise–injection separation metric and the background-consistency metric. The noise–injection separation metric evaluates whether the model preserves the expected sign separation for the glitch and CCSN injections in the validation set. Noise samples without CCSN injections should produce negative outputs, whereas samples containing CCSN injections should yield positive outputs. Maintaining this separation indicates that the model has learned the intended classification behavior. This metric complements the validation cross-entropy, as both evaluate the ability of the model to distinguish injections from noise. However, the background-consistency metric specifically measures whether the model behaves consistently across different background noise conditions. Our goal is for the model to learn generalized glitch features rather than overfitting to noise from H1 or L1 interferometer. Therefore, the average output values for different noise categories should remain as similar as possible. If, in the final training iteration, the average output of any noise category differs from the others by more than $one$, we instead select the checkpoint that minimizes the differences among all noise categories. 

Applying these two metrics to Model\_FF reveals that, despite clear noise–signal separation in the later iterations of Figure~\ref{validation}, the classifier begins to exhibit increasing dependence on specific background types. This indicates the classifier gradually learns the detailed patterns of the glitch, which can reduce overall generalizability. Accordingly, we select the checkpoint at iteration 125 for the test analysis, where the separation performance remains strong while background dependence is minimized. 

\section{Result \label{sec_result}}

\begin{figure}[tb]
    \includegraphics[width=0.85\linewidth]{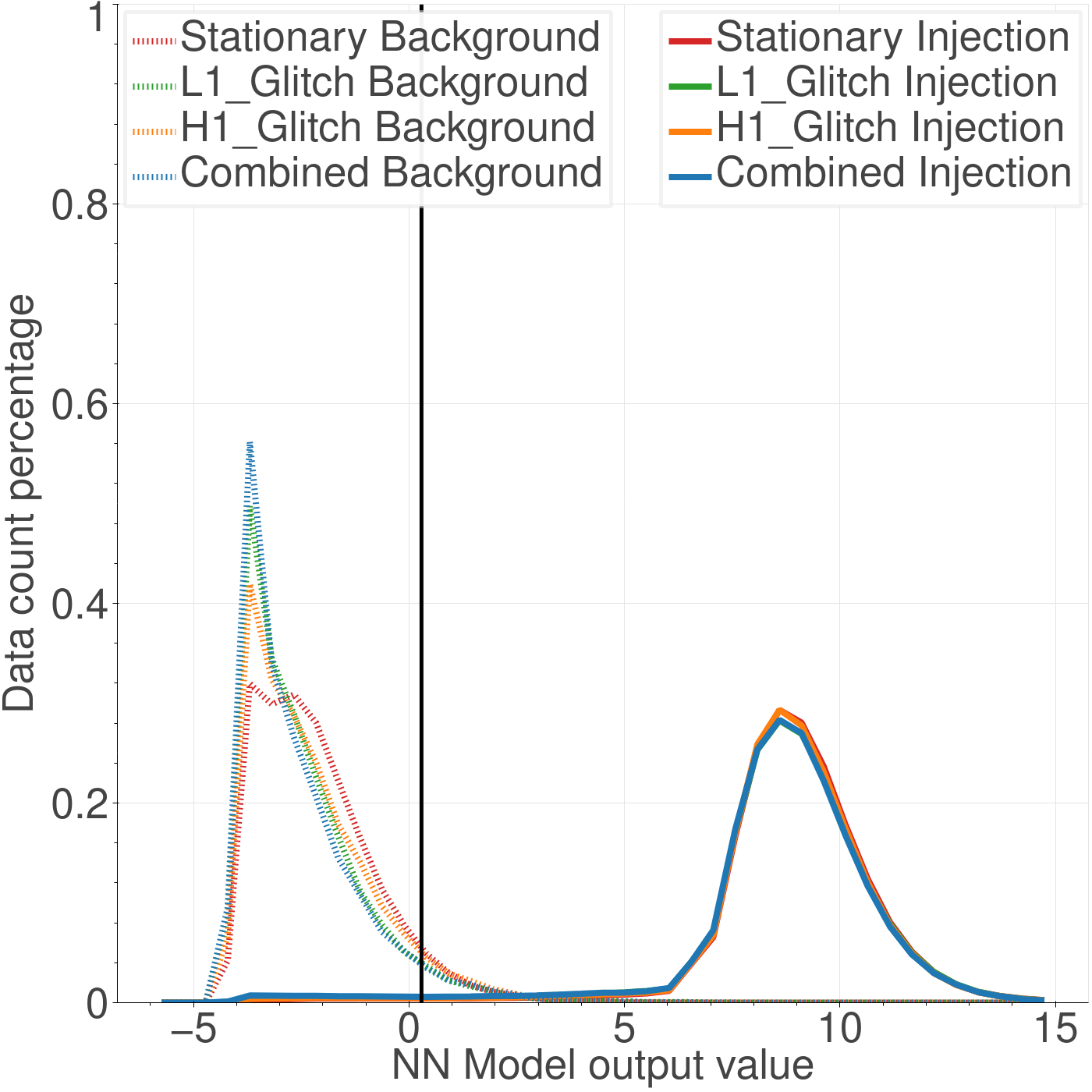}
    \caption{
        Output distribution of noise and CCSN injections of all 31 templates at 20 network SNR from Model\_FF. 
        The solid vertical line indicates the decision threshold separating noise and injection classes. 
    }
    \label{fig:segs_output}
\end{figure}

After training, we pass the combined glitch background to the model to generate an output distribution of the glitch dataset. This distribution is used to determine a decision threshold for distinguishing noise from CCSN-like signals. The threshold is set at a value greater than 95\% of the combined glitch background. Outputs exceeding this threshold are labeled as CCSN events; otherwise, the signal will be labeled as noise-like events. Figure~\ref{fig:segs_output} shows the model output distribution for noise data and CCSN injections. 

We evaluate the sensitivity of our models with CCSNe injections at varying network SNRs and source distances. From these evaluations, we define two metrics: sensitive SNR and sensitive distance. Sensitive SNR is the network SNR at which the model achieves a true positive rate (TPR) of 50\% while maintaining a fixed 5\% false positive rate (FPR) against combined glitches. Here, TPR denotes the fraction of injected signals that exceed the decision threshold. This metric characterizes the signal amplitude relative to the noise level required for detection. Sensitive distance, similarly, is defined as the injection distance corresponding to a 50\% TPR, thereby representing the effective detection range of the model. The false alarm psibility (FAP) is the false postive events devided by the amount of true positive plus false postive events. Under the operating point adopted in this study, corresponding to a TPR of 50\% at a 5\% FPR, the resulting FAP is approximately 10\%. While sensitive SNR measures detection relative to background fluctuations, sensitive distance accounts for astrophysical variability in waveform amplitudes due to progenitor properties and explosion orientation. In the following sections, we first assess the performance of Model\_FF, then in the latter part we compare the other models with Model\_FF to assess the robustness of \texttt{CCSNet}.

\subsection{Full family Full waveform} 

\begin{figure*}[tb]
    \centering
    \begin{subfigure}{0.5\textwidth}
      \centering
      \includegraphics[width=1\linewidth]{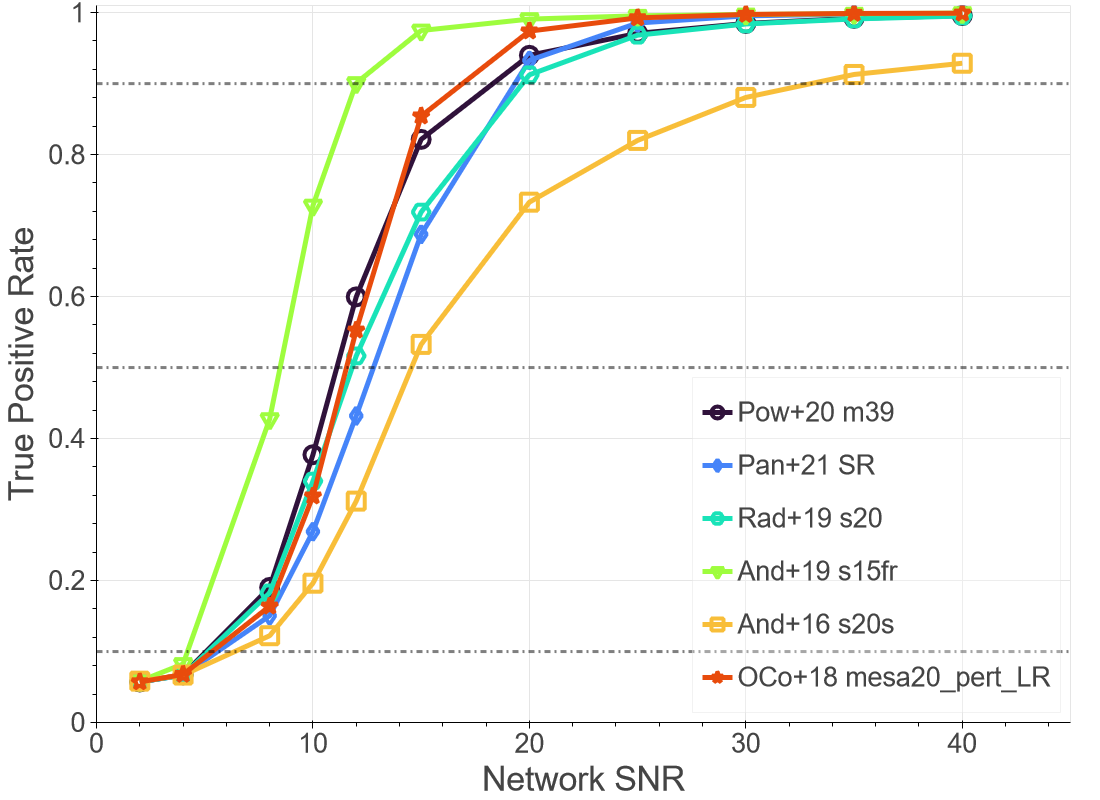} 
    \end{subfigure}%
    \centering
    \begin{subfigure}{0.5\textwidth}
        \centering
        \includegraphics[width=1\linewidth]{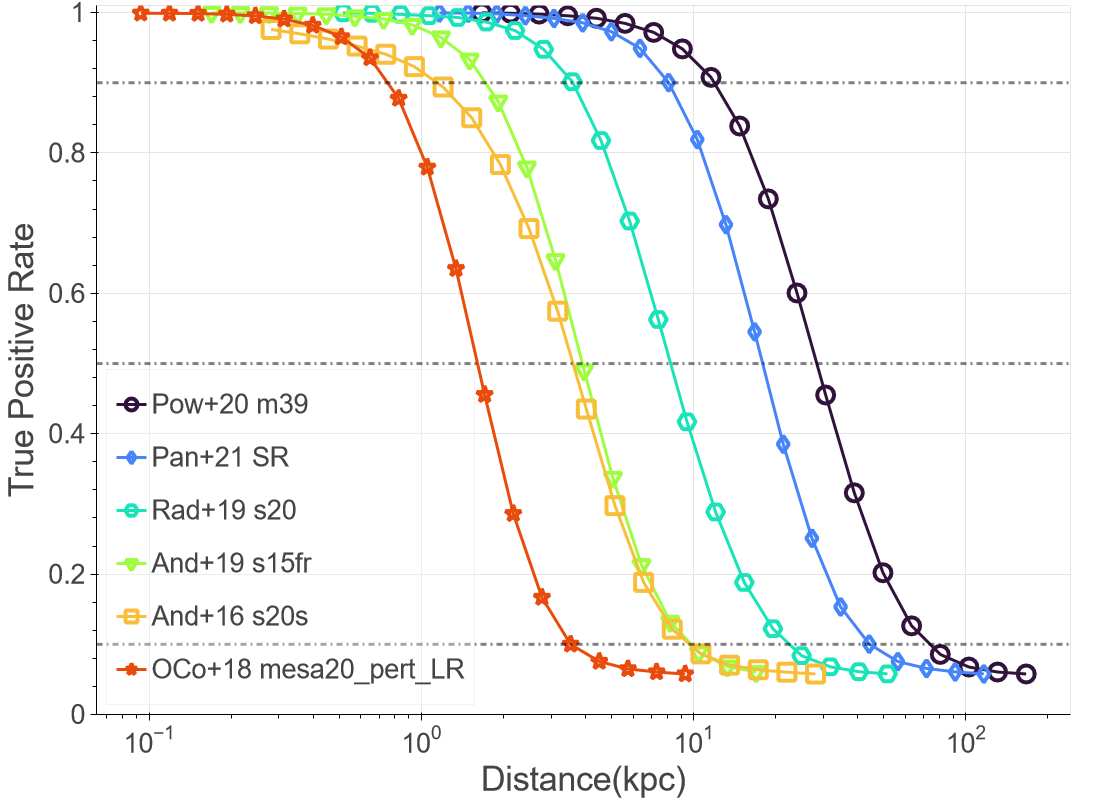} 
    \end{subfigure}%
    \caption{
        Detection performance across different injection intensities (SNR and distance) on stationary backgrounds for the first test segment. 
        The left panel shows TPR versus SNR, and the right panel shows TPR versus injection distance.
    }
    \label{fig:s-curve}
\end{figure*}

\begin{figure*}[tb]
    \centering
    \begin{subfigure}{0.5\textwidth}
        \centering
        \includegraphics[width=1\linewidth]{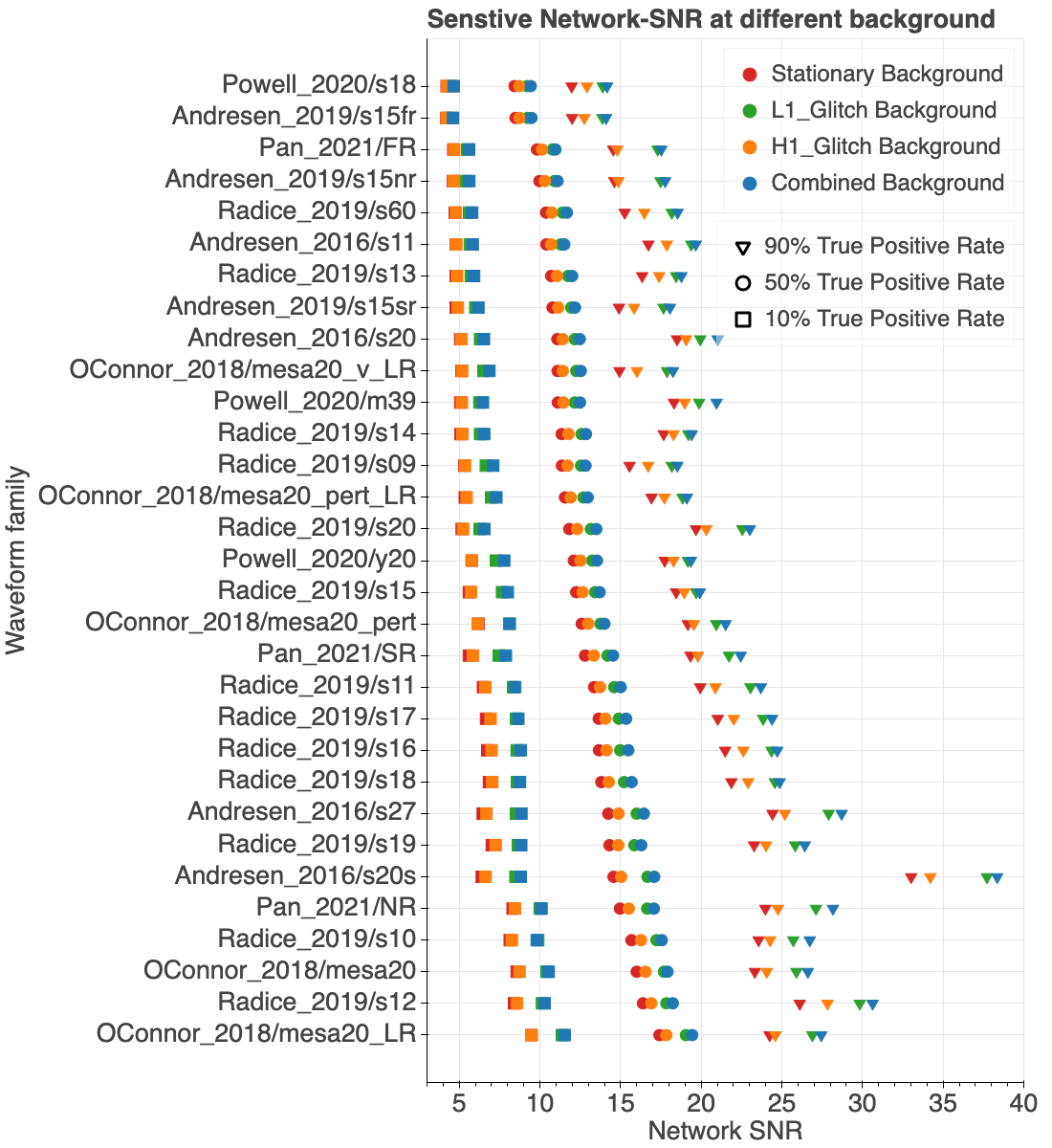} 
    \end{subfigure}%
    \centering
    \begin{subfigure}{0.5\textwidth}
        \centering
        \includegraphics[width=1\linewidth]{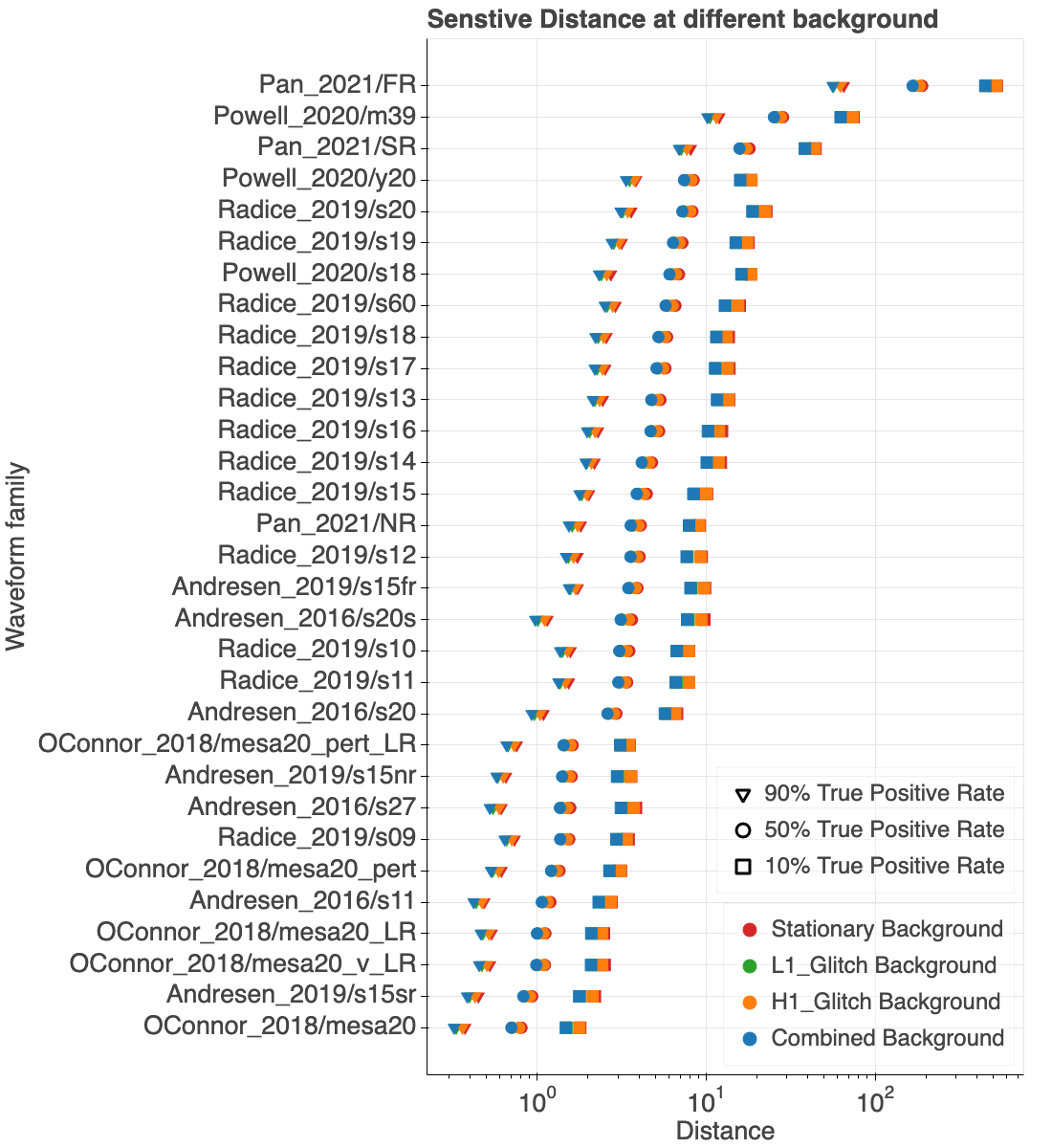} 
    \end{subfigure}%
    \caption{
        Sensitivity SNR (left) and distance (right) for each CCSN template on different background types. The circle dots represent the injection intensity required to achieve a 50\% true positive rate for a given waveform.
    }
    \label{fig:bg-result}
\end{figure*}

Model\_FF is trained on the complete set of CCSN waveforms prepared for this study. Figure \ref{fig:segs_output} presents the output distribution noise data and all 31 CCSNe injections at 20 network SNR, evaluated across all four noise types. To quantify detection performance, we measure the TPR of different CCSN signals at varying injection intensities, as illustrated in Figures \ref{fig:s-curve} and \ref{fig:bg-result}. Figure \ref{fig:s-curve} shows the injection performance on stationary backgrounds for waveforms from different simulation groups. From these curves, we extract the injection intensities corresponding to 90\%, 50\%, and 10\% TPR to construct the summary in Figure \ref{fig:bg-result}, which provides the sensitivity performance of each CCSN injection across all four noise types. Table \ref{tab:performance_table} shows Model\_FF achieves a 50\% recovery rate for normal CCSN injections at a distance of 3.7 kpc under stationary noise. When glitch signals are present in both detectors, the model maintains a sensitivity distance up to 3.5 kpc. For extreme CCSN waveforms, the sensitivity distances reach 77 kpc and 75.6 kpc under stationary and glitch-contaminated backgrounds, respectively. Across both dual- and single-detector configurations, the sensitivity decreases by no more than 7\% in the presence of glitches. This result demonstrates the robustness of Model\_FF in identifying CCSN signals even when the data are corrupted by transient noise artifacts.

\subsection{Model comparison}

\begin{figure*}[tb]
    \centering
    \begin{subfigure}{0.5\textwidth}
    \includegraphics[width=1\linewidth]{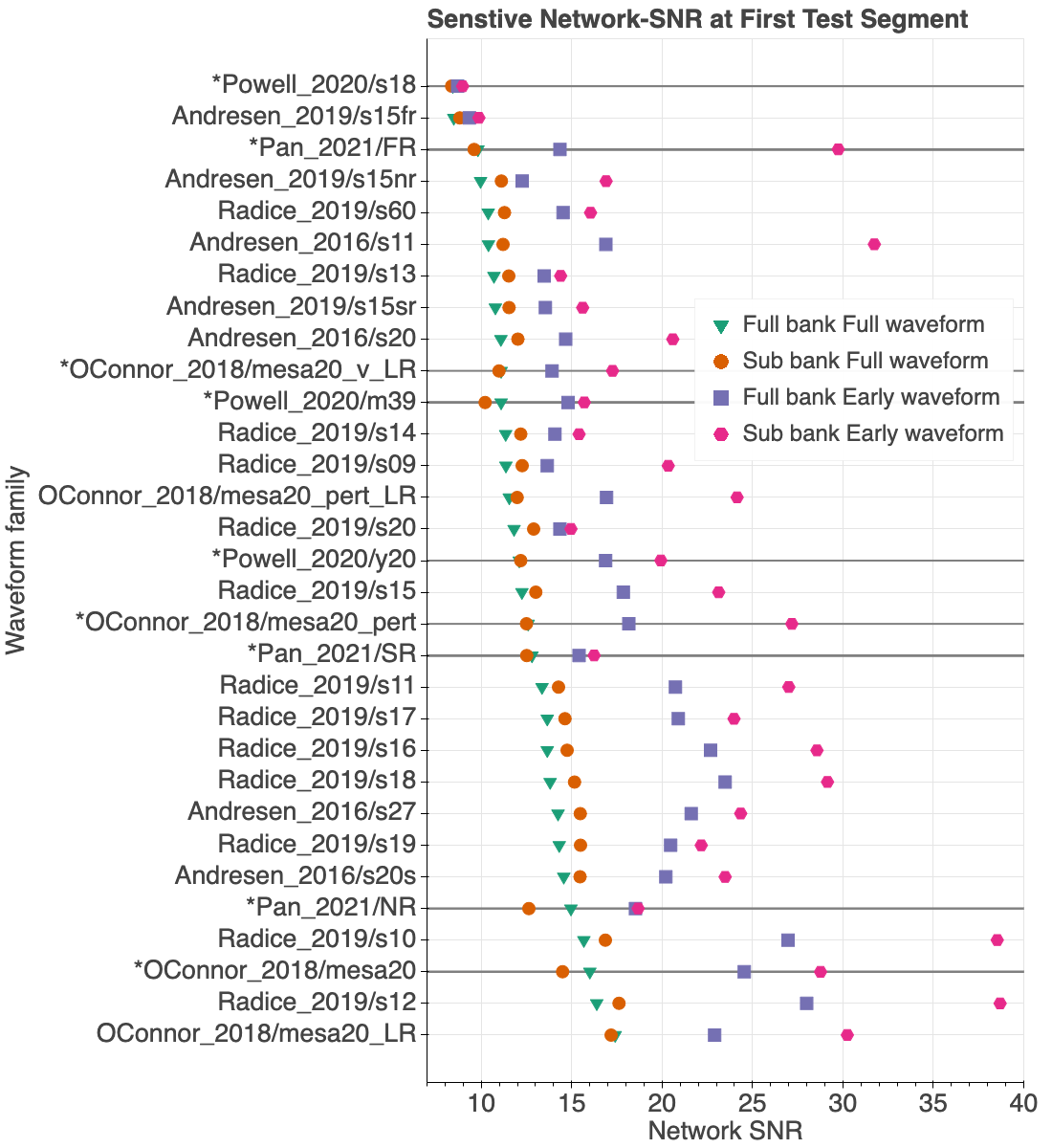}
    \label{fig:method-result-snr}
    \end{subfigure}%
    \begin{subfigure}{0.5\textwidth}
    \includegraphics[width=1\linewidth]{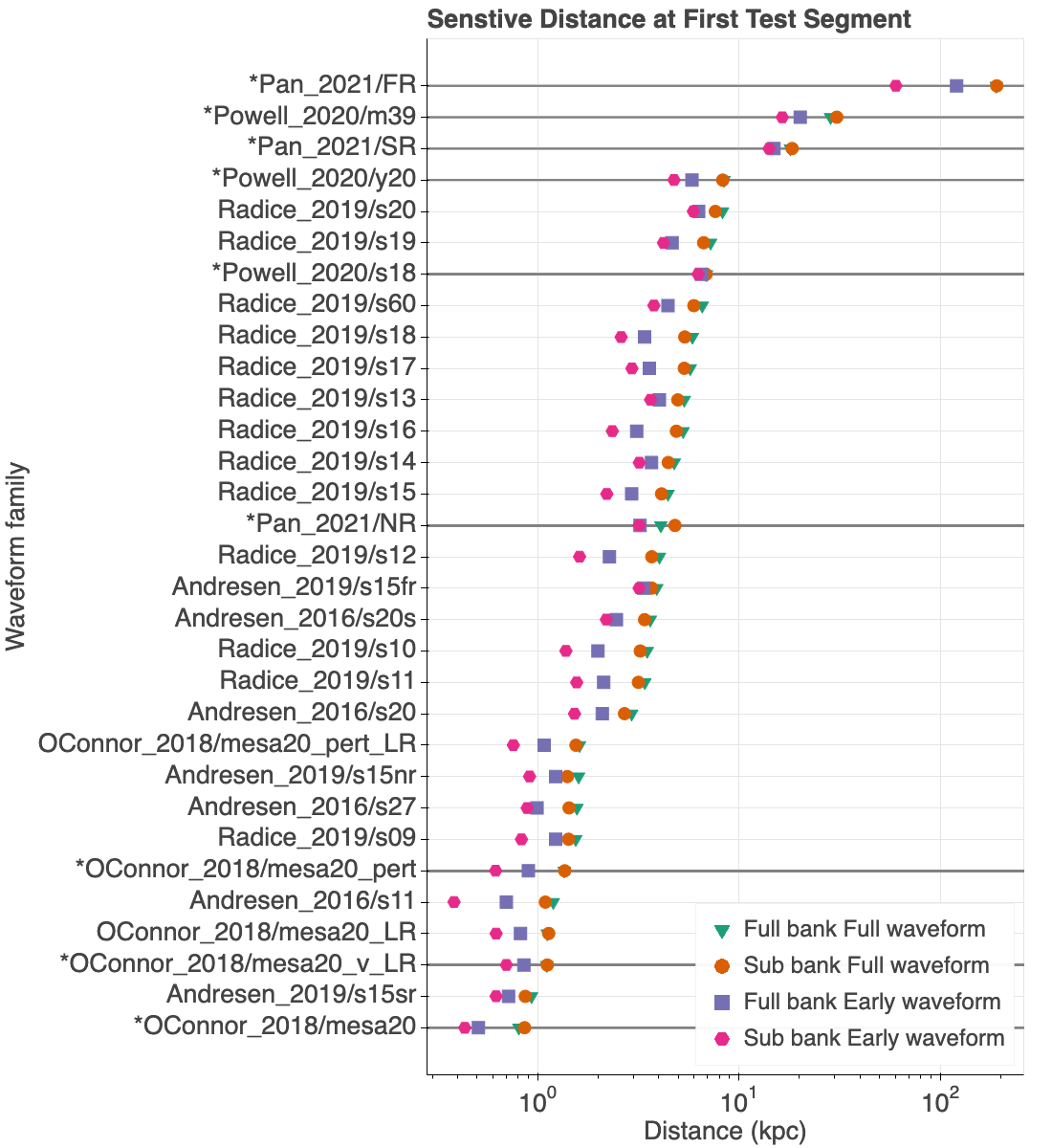}
    \label{fig:method-result-dis}
    \end{subfigure}%
    \caption{
        Sensitivity comparison of different training methods evaluated on the stationary background. 
        The left panel shows the sensitivity SNR, and the right panel shows the sensitivity distance. 
        Waveform families marked with an asterisk are the subgroup CCSN waveforms.
    }
    \label{fig:method-result}
\end{figure*}

Using Model\_FF as our baseline model, we compare the sensitivity of three alternative models: Model\_SF (sub-template bank, full waveform), Model\_FE (full template bank, early waveform), and Model\_SE (sub-template bank, early waveform). Figure~\ref{fig:method-result} shows that Model\_SF achieves sensitivity comparable to Model\_FF across the full test set, indicating that the model is capable of capturing CCSN templates that were not included in training. In contrast, Model\_FE and Model\_SE are trained specifically on early-phase CCSN waveforms. These models retain up to 65\% and 40\% of the sensitivity distance for normal and extreme waveforms, respectively. Although they are trained using truncated signals of approximately 100 milliseconds, the loss of  sensitivity distance are nealy negligible when glitch signals appear in the input coincidentally. These results highlight the capability of \texttt{CCSNet} to generate triggers selectively from early-phase CCSN signals while maintaining robustness against transient noise artifacts.

\begin{figure*}[tb]
    \centering
    \begin{subfigure}{0.5\textwidth}
        \centering
        \includegraphics[width=\linewidth]{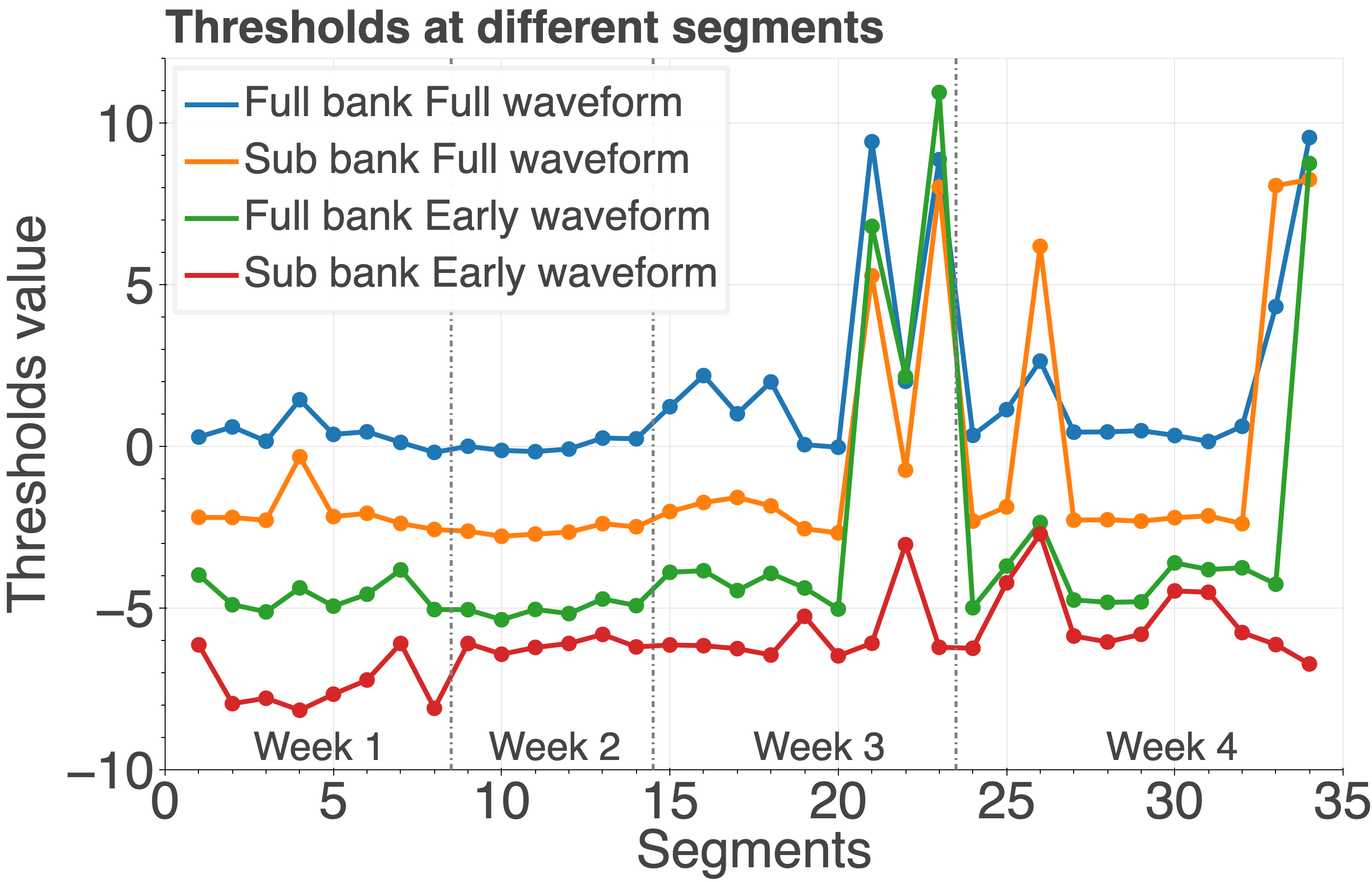}
    \end{subfigure}%
    \begin{subfigure}{0.5\textwidth}
        \centering
        \includegraphics[width=\linewidth]{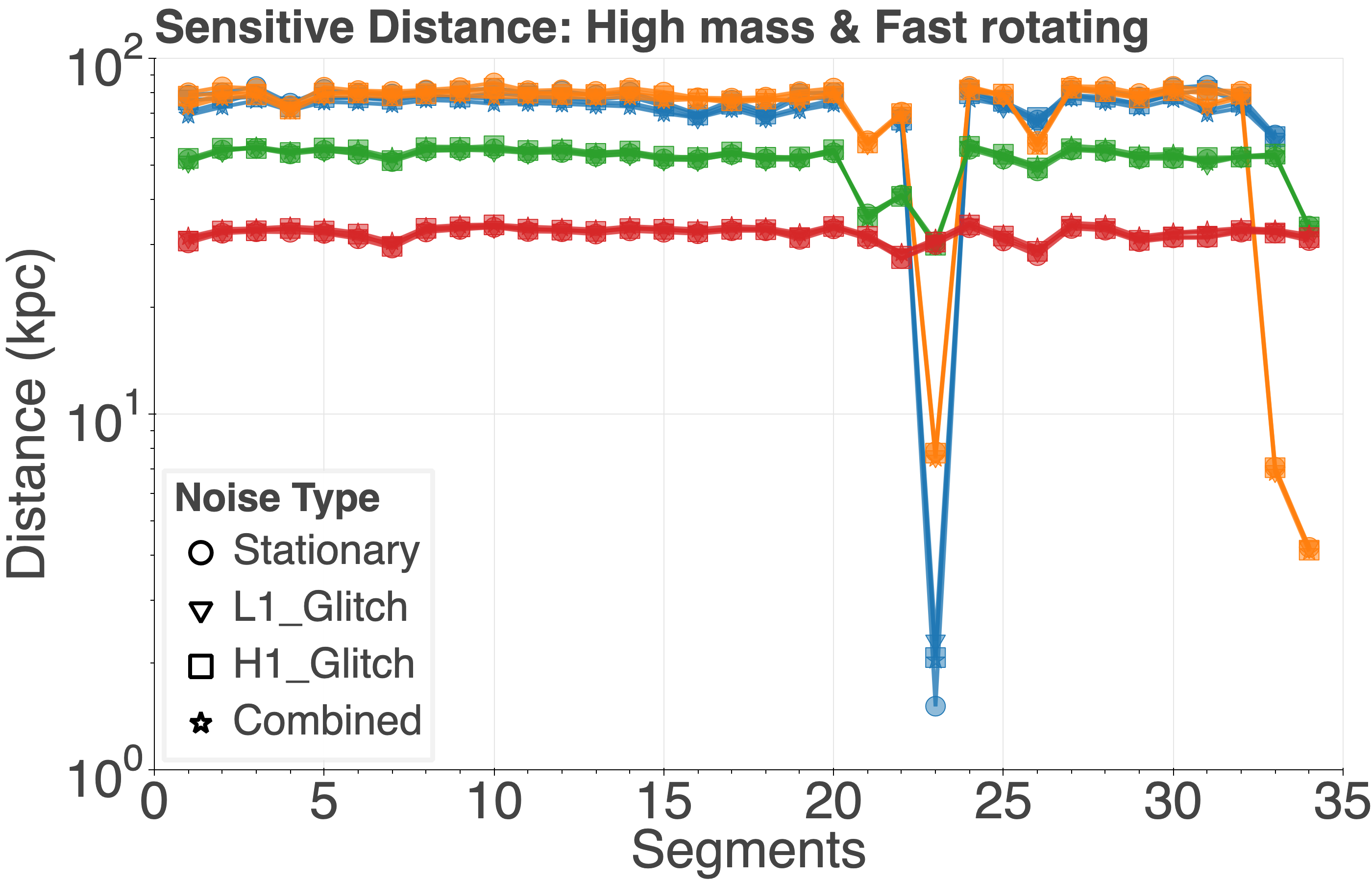}
    \end{subfigure}%

    \begin{subfigure}{0.5\textwidth}
        \centering
        \includegraphics[width=\linewidth]{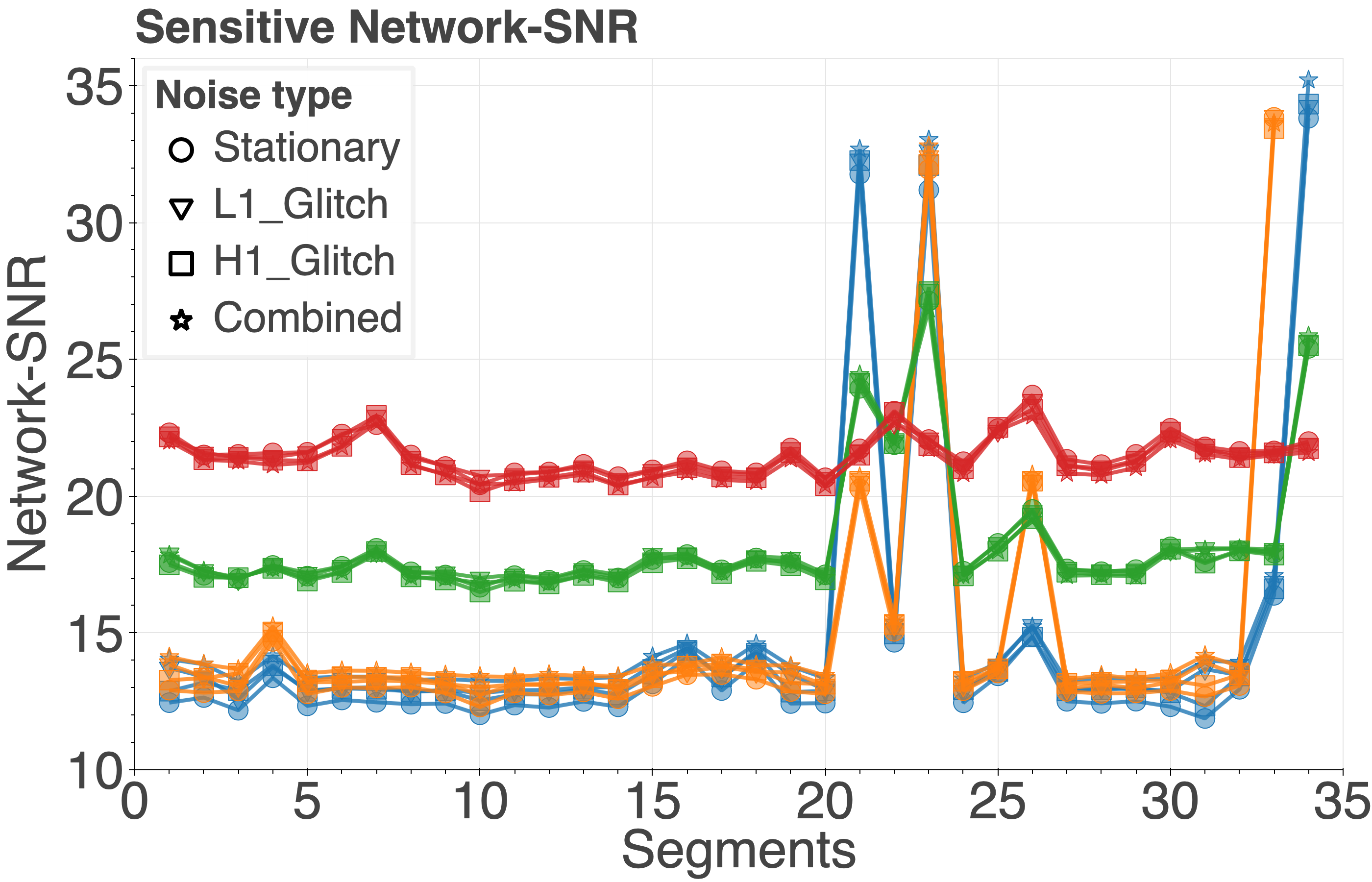}
    \end{subfigure}%
    \begin{subfigure}{0.5\textwidth}
        \centering
        \includegraphics[width=\linewidth]{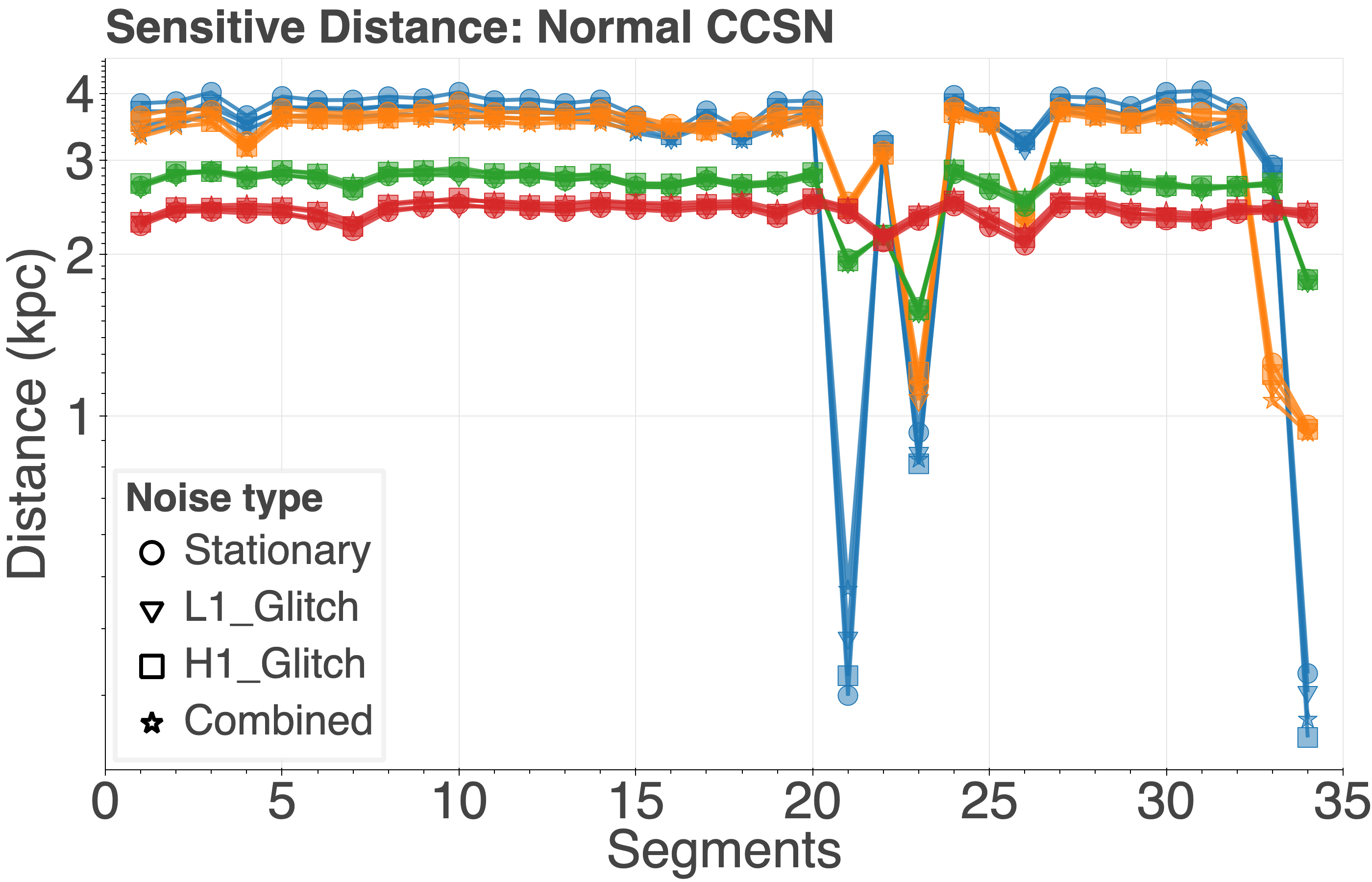}
    \end{subfigure}%
    \caption{Segment-by-segment performance evaluation across four weeks of strain data for the four models. 
        The top-left panel shows the evolution of decision thresholds for the four models. 
        Thresholds remain stable for most segments, except for segments 21, 23, 33, and 34, indicating optimization or retrain is needed. The bottom-left panel presents the corresponding sensitive network SNR, showing that Model\_FF and Model\_SF achieve nearly identical performance throughout the dataset. The top-right and bottom-right panels display the sensitive distances for high-mass, fast-rotating CCSNe and for normal CCSNe, respectively. Overall, the results show that all four models maintain perform similary on stationary and glitch-contaminated backgrounds, indicating their ability to detect CCSN signals even when overlapping with glitches.}
    \label{fig:longevity_performance}
\end{figure*}

To examine long-term stability, we evaluate all models on data recorded four weeks after the training period. Since the CCSN templates span a wide range of physical properties and signal amplitudes, we classify them into two groups for interpretation of the results. The first group consists of highly energetic CCSNe originating from massive progenitors with extreme rotation or strong magnetic fields, such as Pan\_FR, Pan\_SR, and Powell\_m39, which exhibit amplitudes one to two orders of magnitude larger than the others at the same distance. The second group includes the remaining CCSN templates, which display more moderate physical properties.The average sensitive SNR and sensitive distance across these segments are summarized in Figure~\ref{fig:longevity_performance}, with additional details provided in Table~\ref{tab:performance_table}. 

\begin{figure*}[tb]
    \centering
    \begin{subfigure}{0.25\textwidth}
        \centering
        \includegraphics[width=\linewidth]{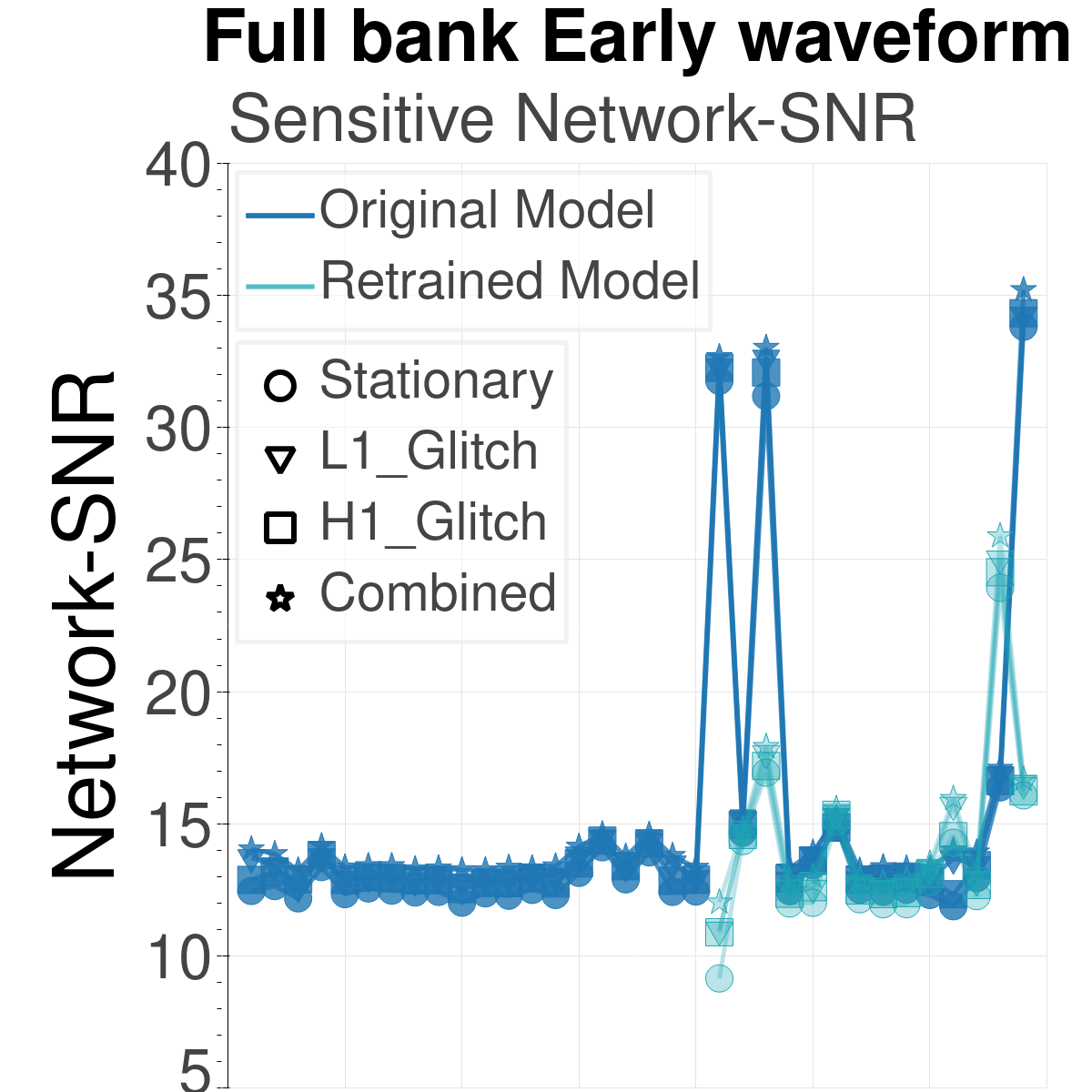}
    \end{subfigure}%
    \begin{subfigure}{0.25\textwidth}
        \includegraphics[width=\linewidth]{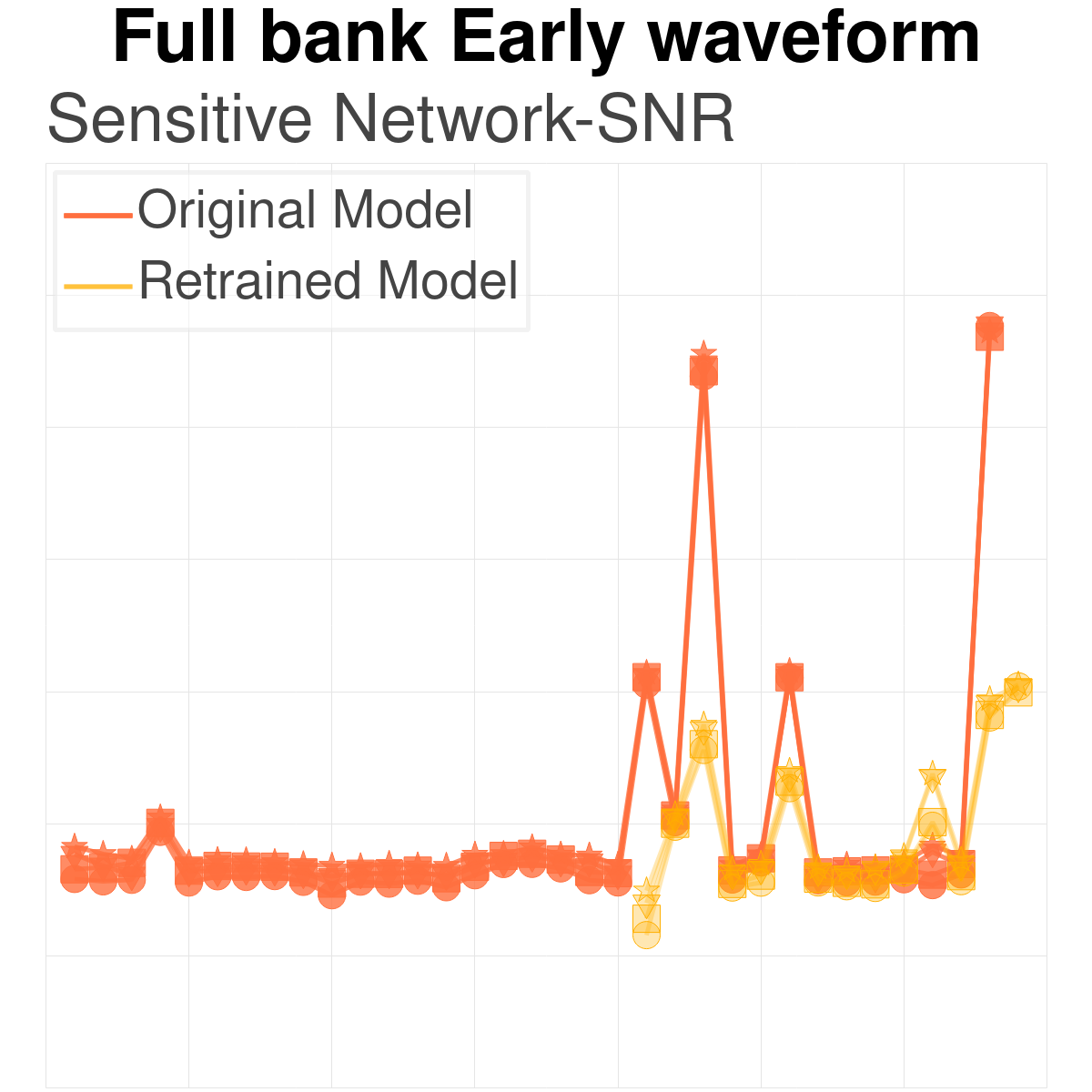}
    \end{subfigure}%
    \begin{subfigure}{0.25\textwidth}
        \includegraphics[width=\linewidth]{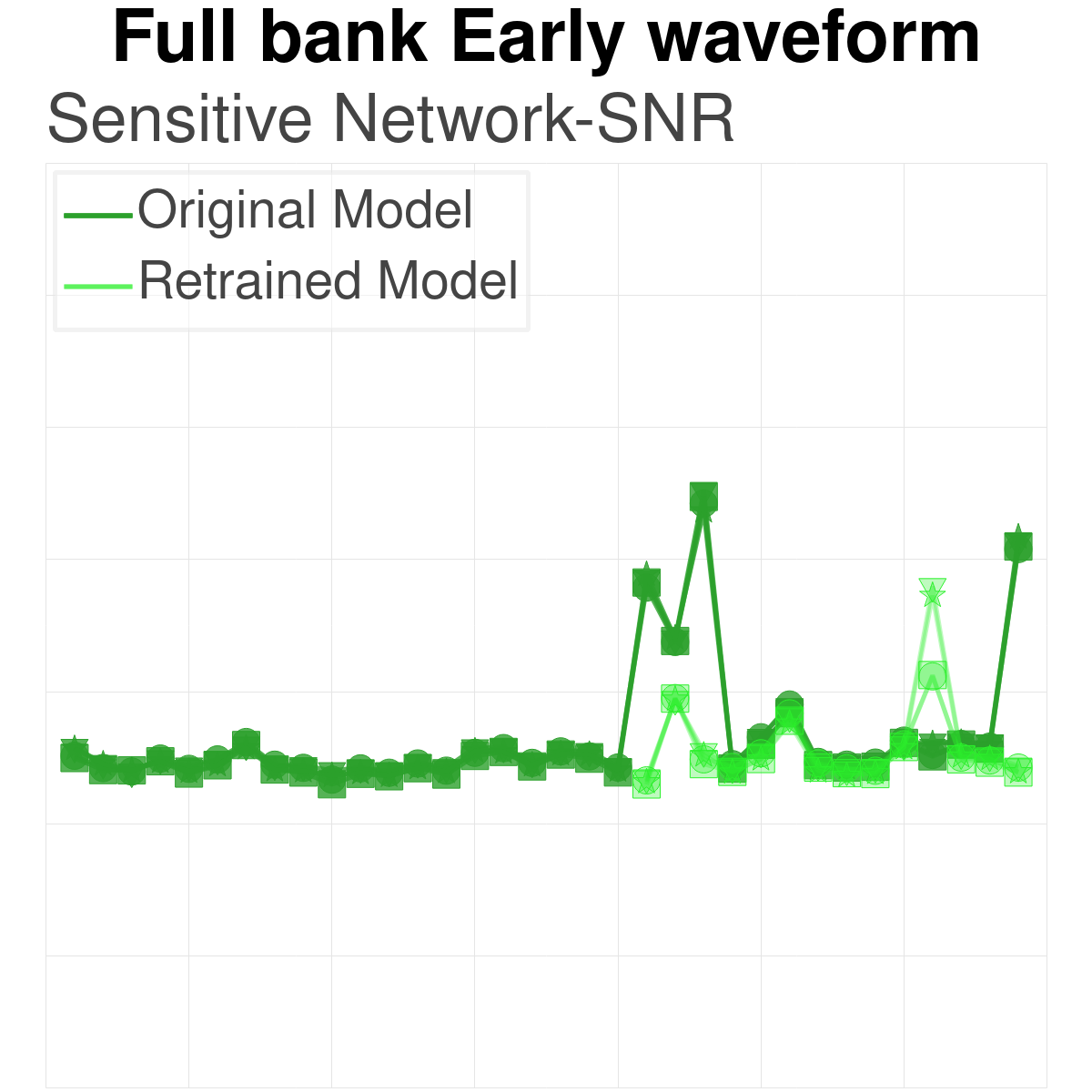}
    \end{subfigure}%
    \begin{subfigure}{0.25\textwidth}
        \includegraphics[width=\linewidth]{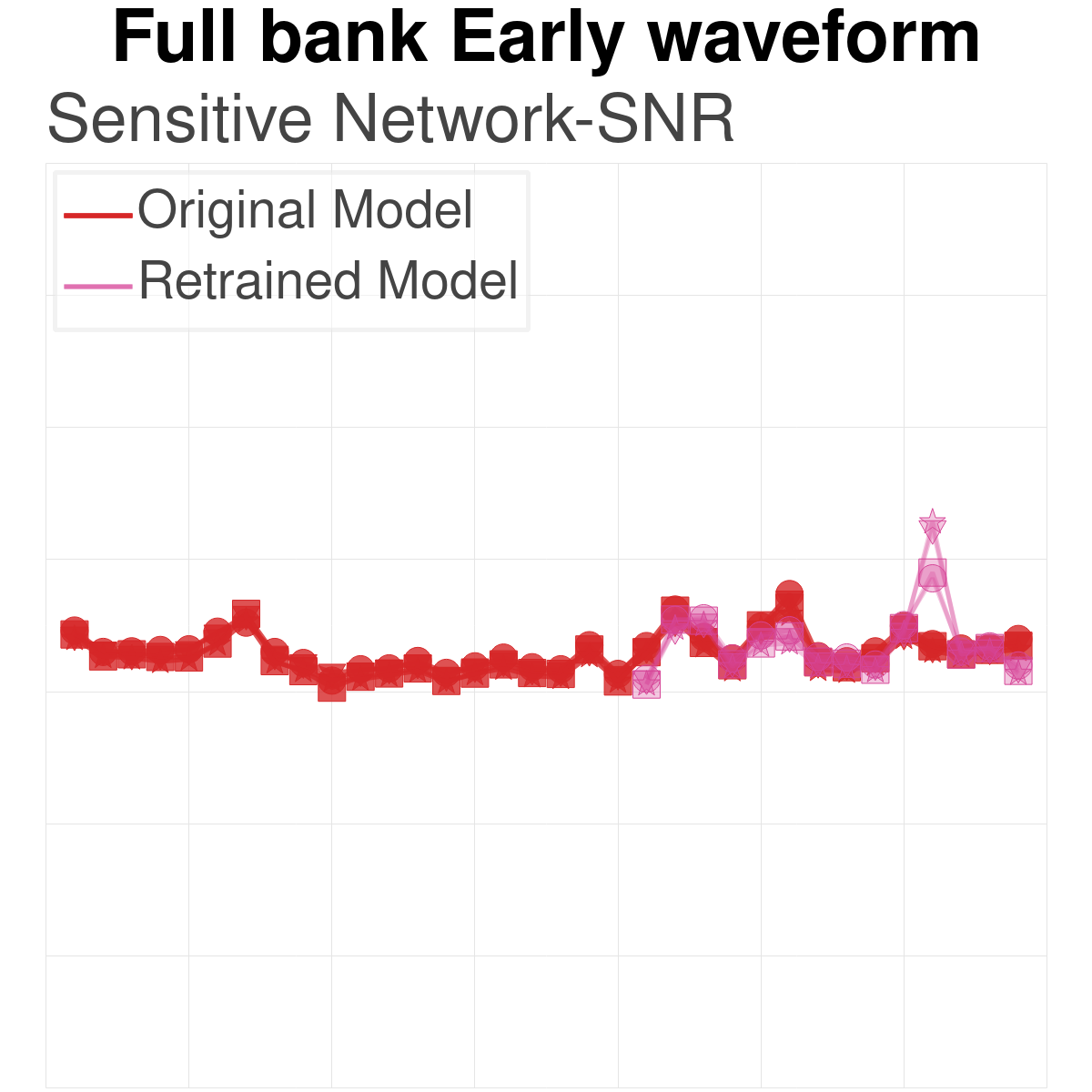}
    \end{subfigure}%

    \begin{subfigure}{0.25\textwidth}
        \centering
        \includegraphics[width=\linewidth]{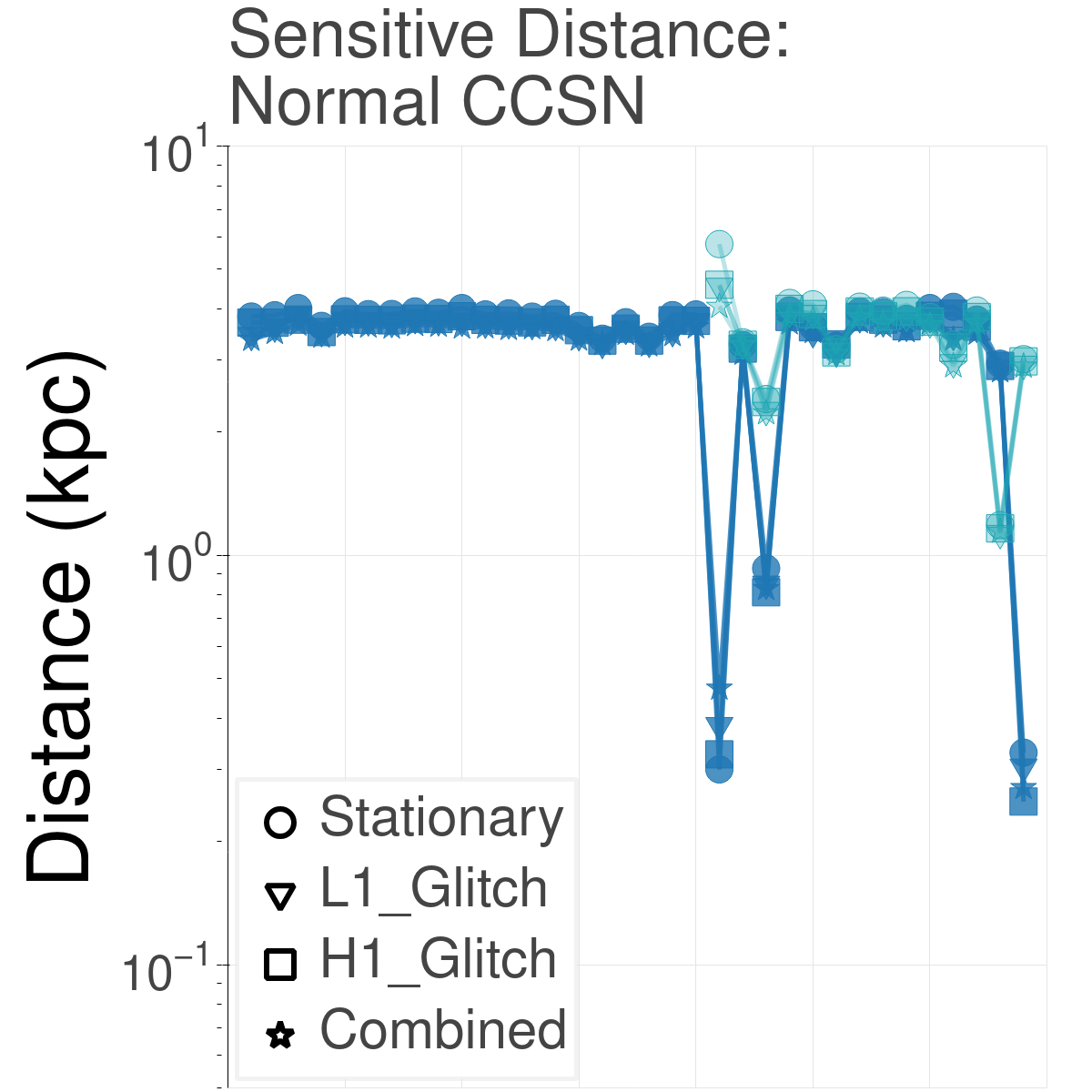}
    \end{subfigure}%
    \begin{subfigure}{0.25\textwidth}
        \includegraphics[width=\linewidth]{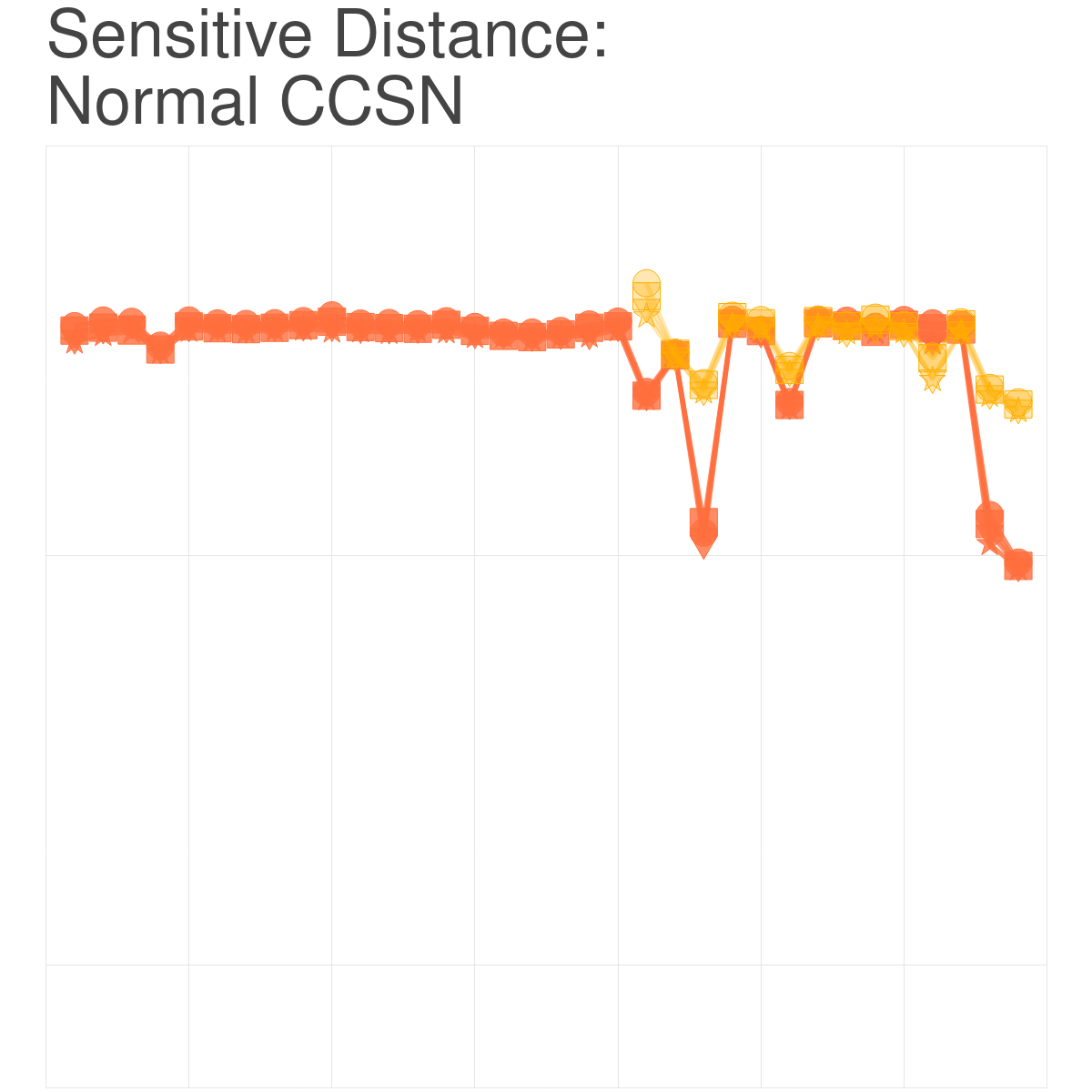}
    \end{subfigure}%
    \begin{subfigure}{0.25\textwidth}
        \includegraphics[width=\linewidth]{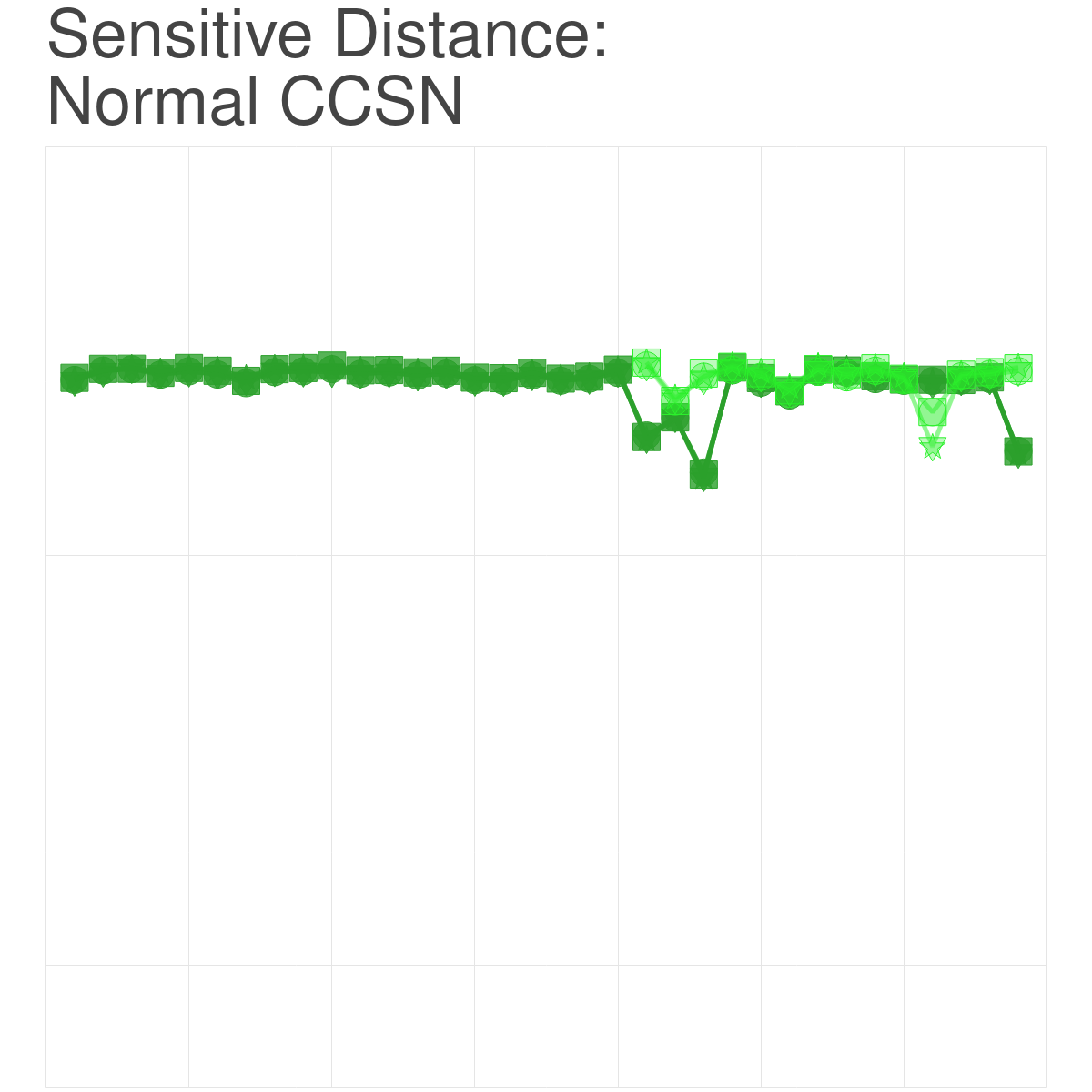}
    \end{subfigure}%
    \begin{subfigure}{0.25\textwidth}
        \centering
        \includegraphics[width=\linewidth]{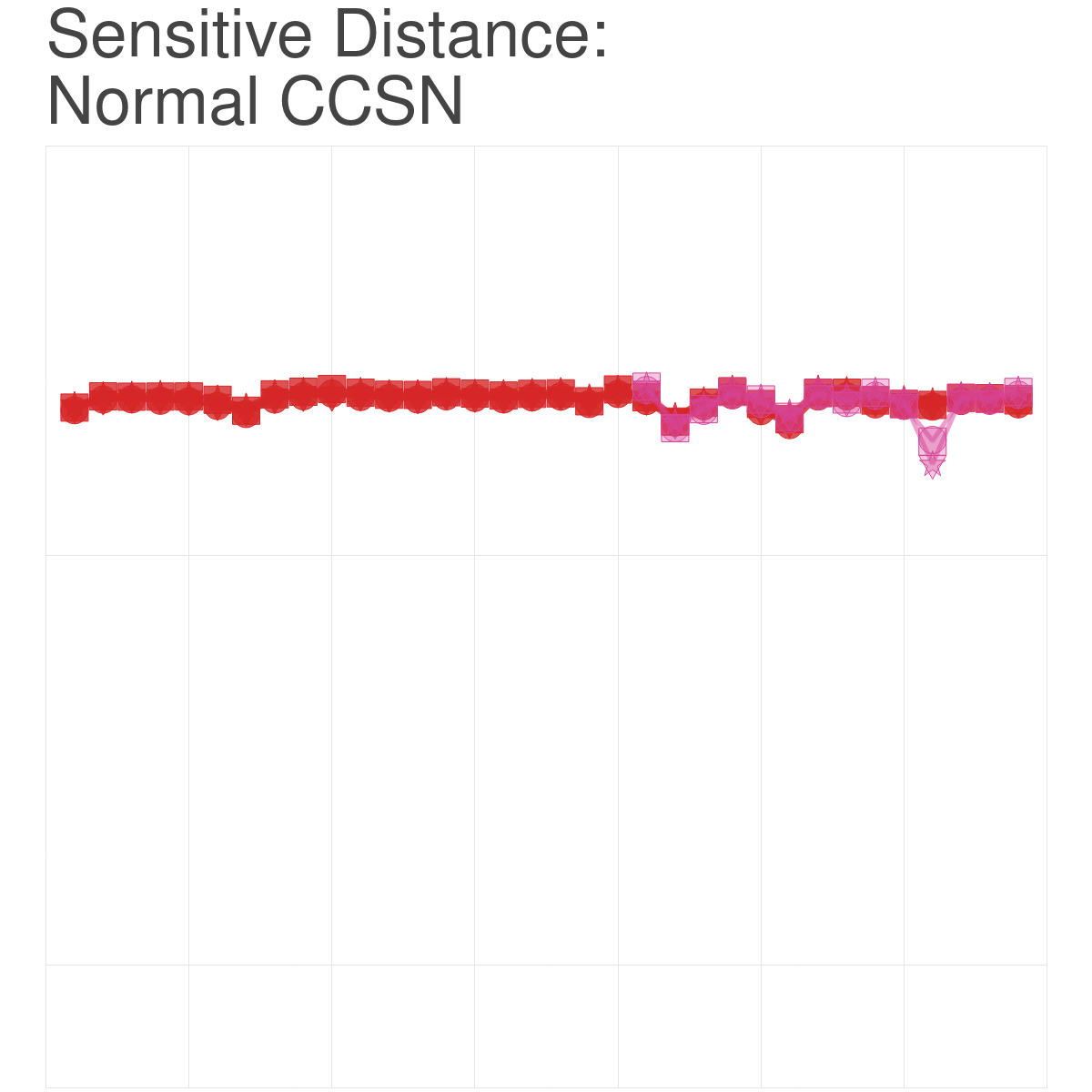}
    \end{subfigure}%
    
    \begin{subfigure}{0.25\textwidth}
        \centering
        \includegraphics[width=\linewidth]{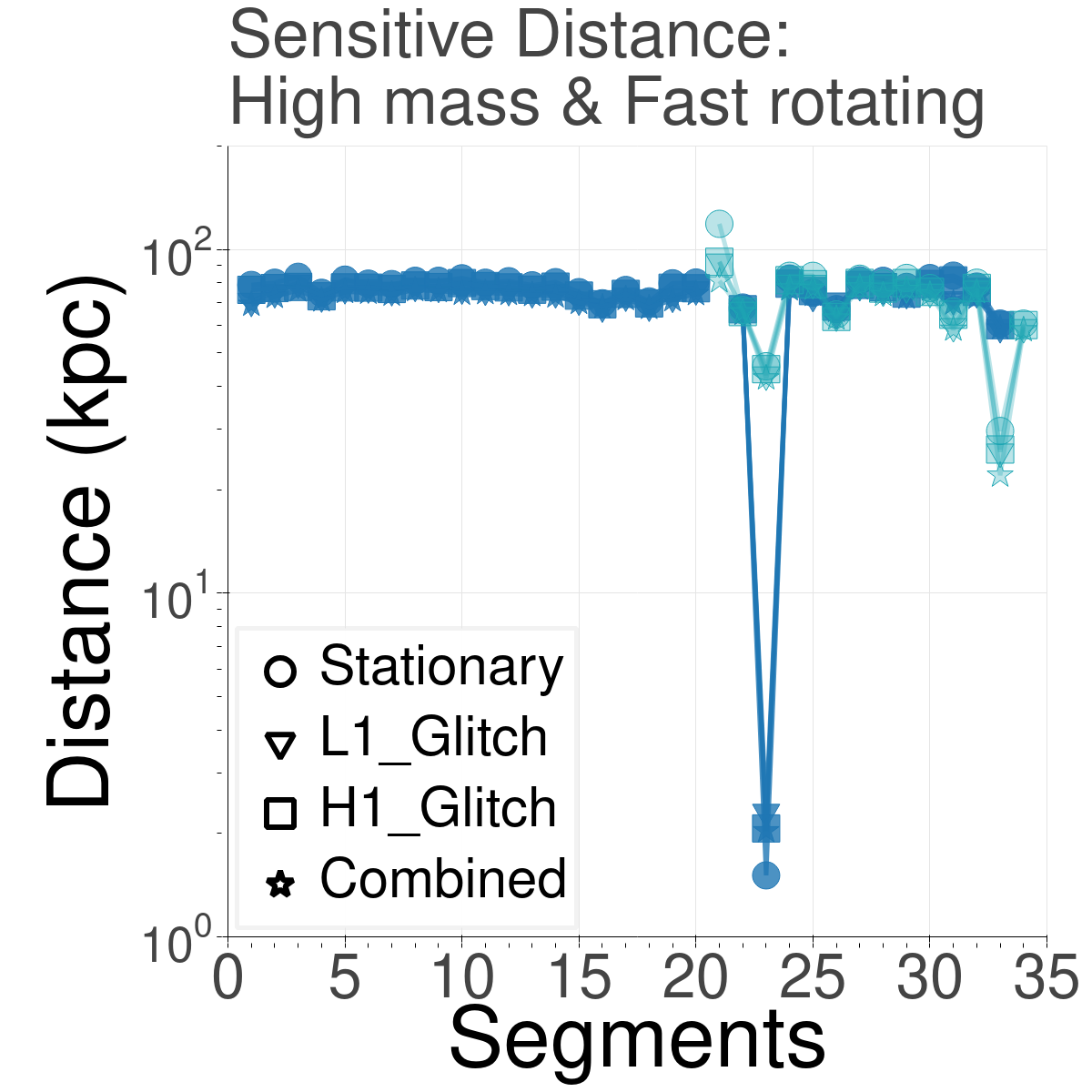}
    \end{subfigure}%
    \begin{subfigure}{0.25\textwidth}
        \includegraphics[width=\linewidth]{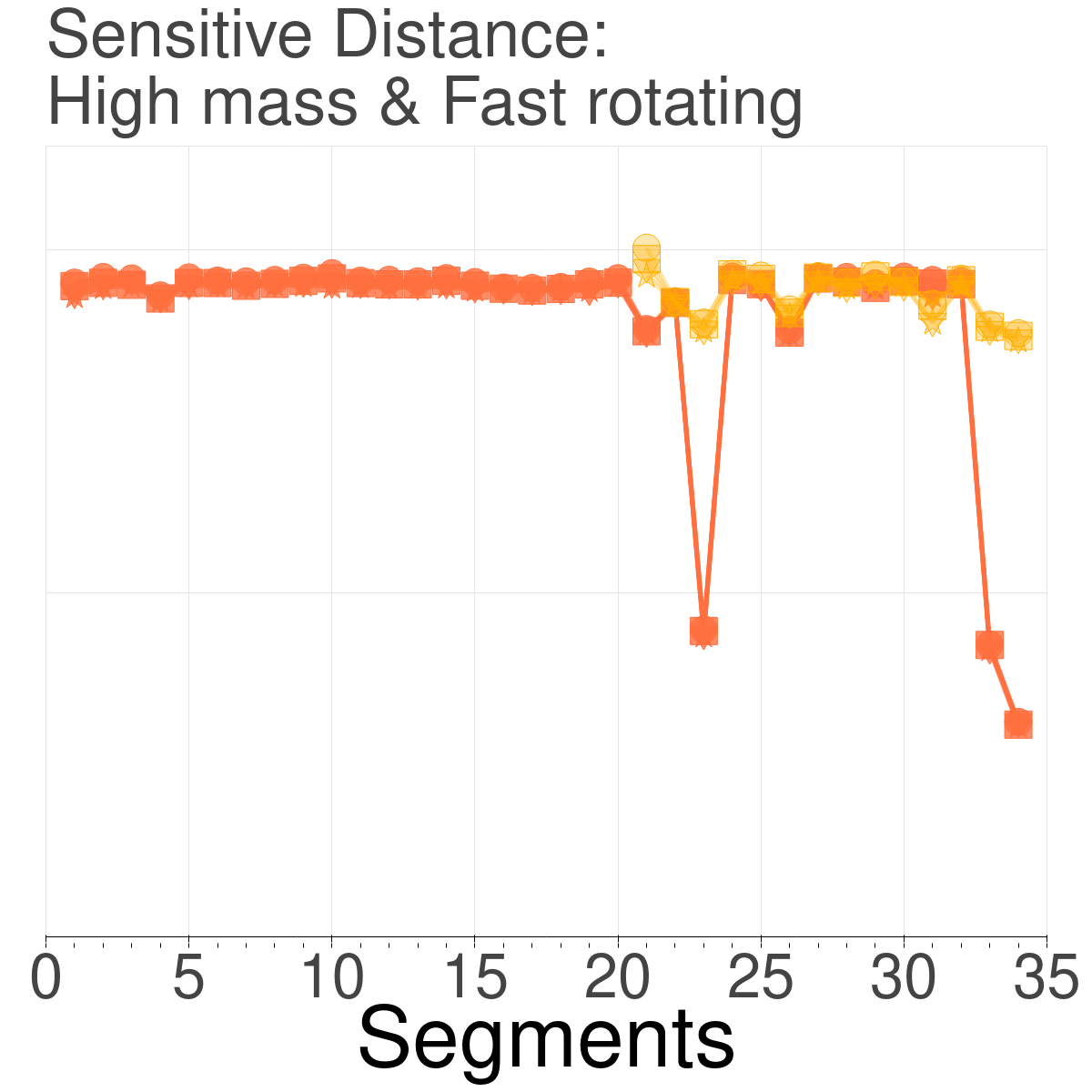}
    \end{subfigure}%
    \begin{subfigure}{0.25\textwidth}
        \includegraphics[width=\linewidth]{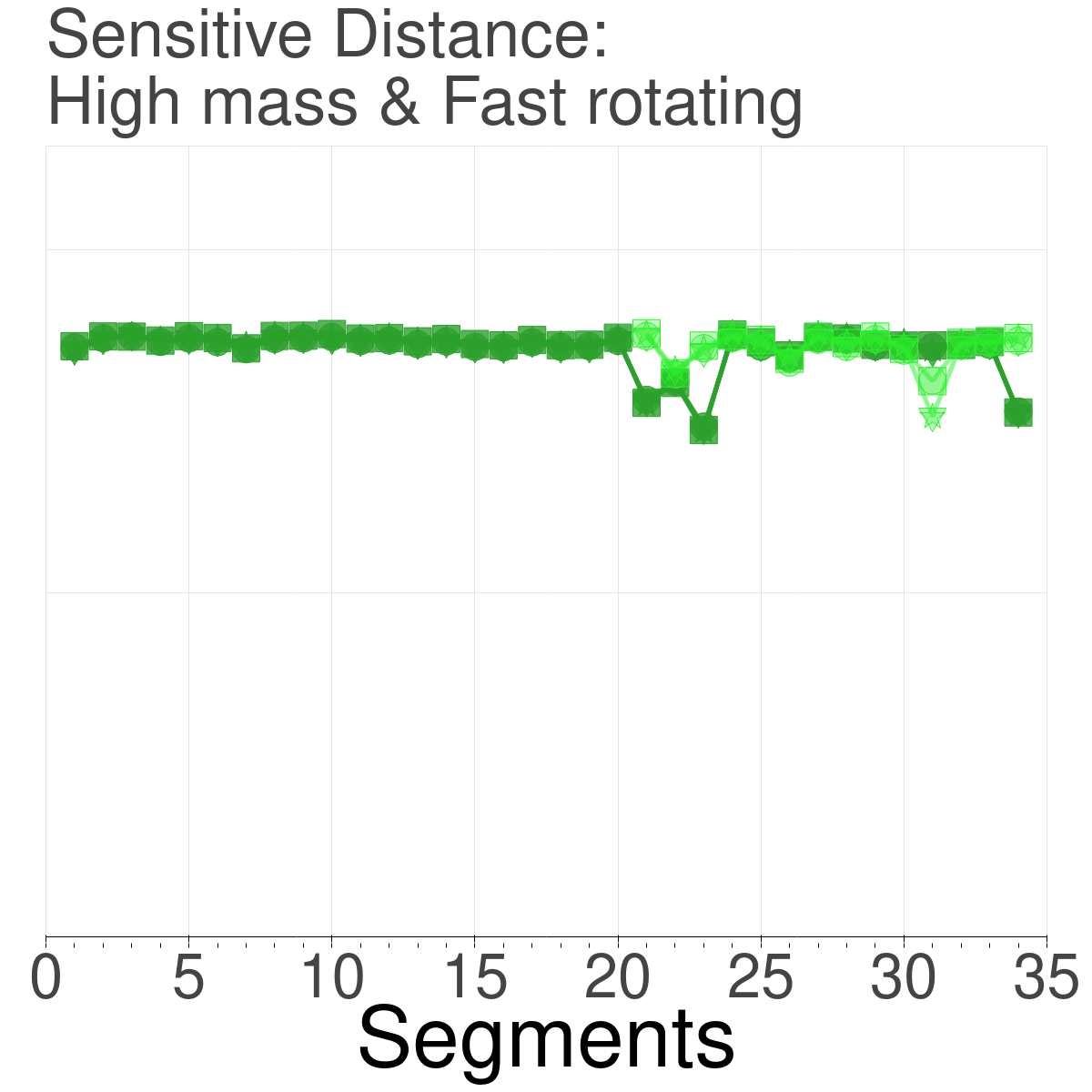}
    \end{subfigure}%
    \begin{subfigure}{0.25\textwidth}
        \centering
        \includegraphics[width=\linewidth]{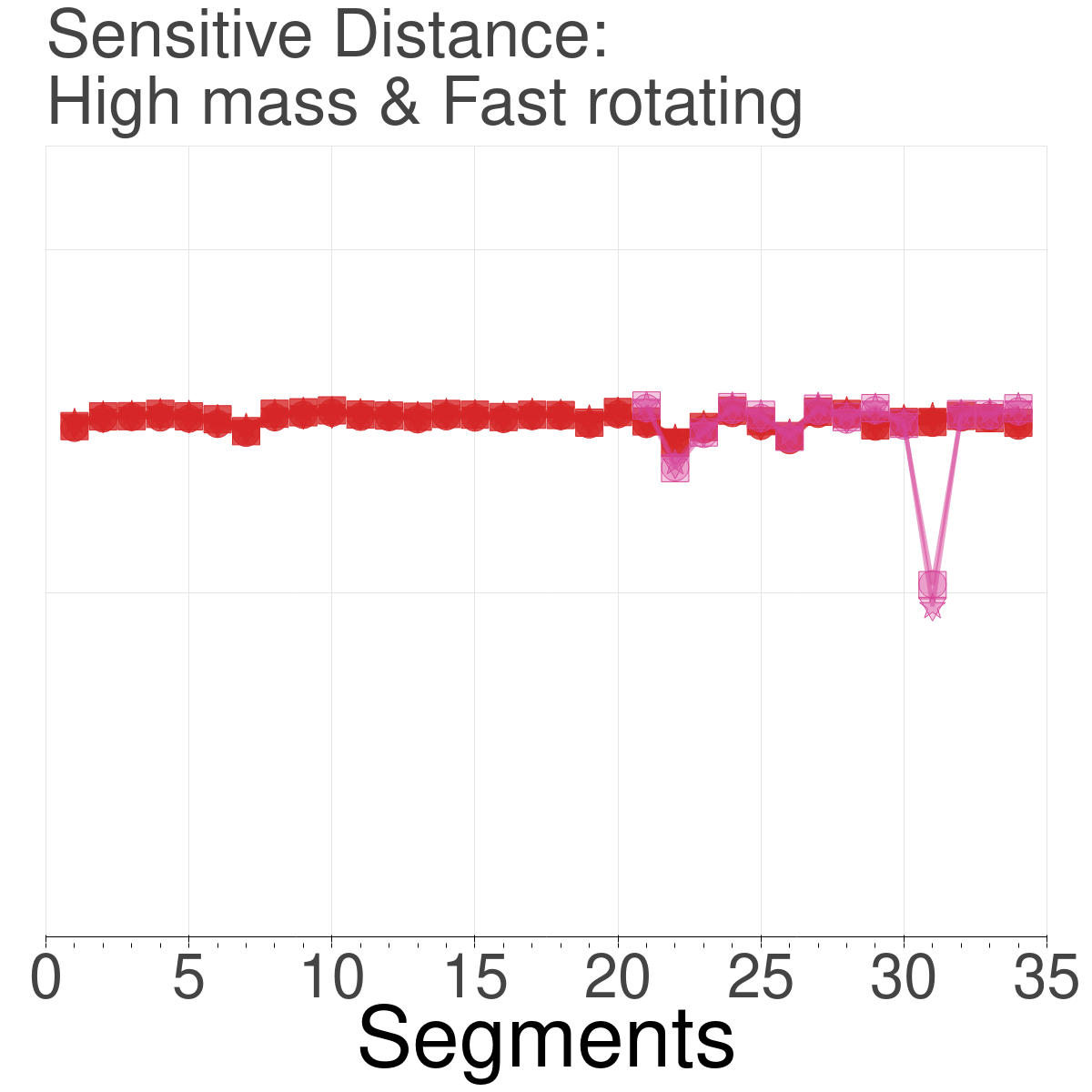}
    \end{subfigure}%
    \caption{
Comparison of sensitivity performance between the original and retrained models across 34 test segments. The top row compares the sensitive network SNR of each model, the middle row corresponds to the sensitive distance for normal CCSN, and the bottom row shows the sensitive distance for extreme CCSN. Each plot compares the original model (solid lines) and the retrained model (faint lines), evaluated on four types of background: stationary, L1 glitch, H1 glitch, and combined. Retraining is performed on segment 21, which is the first segment showing notable performance drops in the original model. Results show that retraining helps restore sensitivity in degraded segments, particularly for the Full waveform configurations.
}
    \label{fig:segs_performance_retrain}
\end{figure*}

Overall, the sensitivity of the models remains stable across most segments. In cases where performance degradation occurs, which can be discovered when the threshold shifts drastically as demonstrated in Figure~\ref{fig:segs_performance_retrain}, can recover the sensitivity by retraining.
These results indicate that targeted retraining offers an effective approach for maintaining performance under evolving detector noise conditions.

\begin{table*}[tb]
\begin{tabularx}{\linewidth}{ccc Y Y Y} 
\hline\hline 
 Injection Method & Noise Type & Detector & 
 Sensitive SNR & 
 \begin{tabular}[c]{@{}c@{}} Normal CCSN \\ Sensitive Distance (kpc) \end{tabular} & 
 \begin{tabular}[c]{@{}c@{}} Extreme CCSN \\ Sensitive Distance (kpc) \end{tabular} \\
\hline
Full template Full waveform     & Stationary & Dual     & 13.5 & 3.7 & 77.0 \\
                                &            & H1       & 18.6 & 2.4 & 49.7 \\
                                &            & L1       & 15.7 & 3.1 & 63.1 \\
                                & Glitch     & Dual     & 14.4 & 3.5 & 71.7 \\
                                &            & H1       & 18.7 & 2.4 & 49.8 \\
                                &            & L1       & 16.3 & 3.0 & 61.1 \\
\hline
Sub template Full waveform      & Stationary & Dual     & 14.0 & 3.5 & 75.6 \\
                                &            & H1       & 18.4 & 2.5 & 56.5 \\
                                &            & L1       & 16.8 & 2.8 & 66.7 \\
                                & Glitch     & Dual     & 14.6 & 3.4 & 72.3 \\
                                &            & H1       & 18.4 & 2.5 & 57.2 \\
                                &            & L1       & 17.4 & 2.7 & 65.5 \\
\hline
Full template Early waveform    & Stationary & Dual     & 18.3 & 2.6 & 51.3 \\
                                &            & H1       & 22.1 & 2.2 & 41.8 \\
                                &            & L1       & 22.7 & 2.1 & 26.1 \\
                                & Glitch     & Dual     & 18.2 & 2.7 & 52.1 \\
                                &            & H1       & 22.1 & 2.2 & 42.3 \\
                                &            & L1       & 23.1 & 2.0 & 24.8 \\
\hline
Sub template Early waveform     & Stationary & Dual     & 21.6 & 2.4 & 31.7 \\
                                &            & H1       & 25.4 & 1.8 & 17.6 \\
                                &            & L1       & 26.1 & 1.9 & 16.1 \\
                                & Glitch     & Dual     & 21.3 & 2.4 & 32.8 \\
                                &            & H1       & 25.5 & 1.9 & 17.8 \\
                                &            & L1       & 26.1 & 1.9 & 16.1 \\
\hline\hline
\end{tabularx}
\caption{
    Sensitivity performance of CCSNet models trained with different injection strategies (full vs. sub template) and waveform durations (full vs. early waveform), evaluated under stationary and glitch-contaminated noise backgrounds. Results are reported for dual-detector (H1–L1) and single-detector (H1 or L1) configurations. For each setup, we list the sensitive SNR threshold and the sensitive distances (in kpc) for both normal and extreme CCSN waveforms. All sensitivity values are averaged over the complete test set of 31 CCSN signals.
}
\label{tab:performance_table}
\end{table*}

\begin{figure*}[tb]
    \centering
    \begin{subfigure}{0.25\textwidth}
        \centering
        \includegraphics[width=\linewidth]{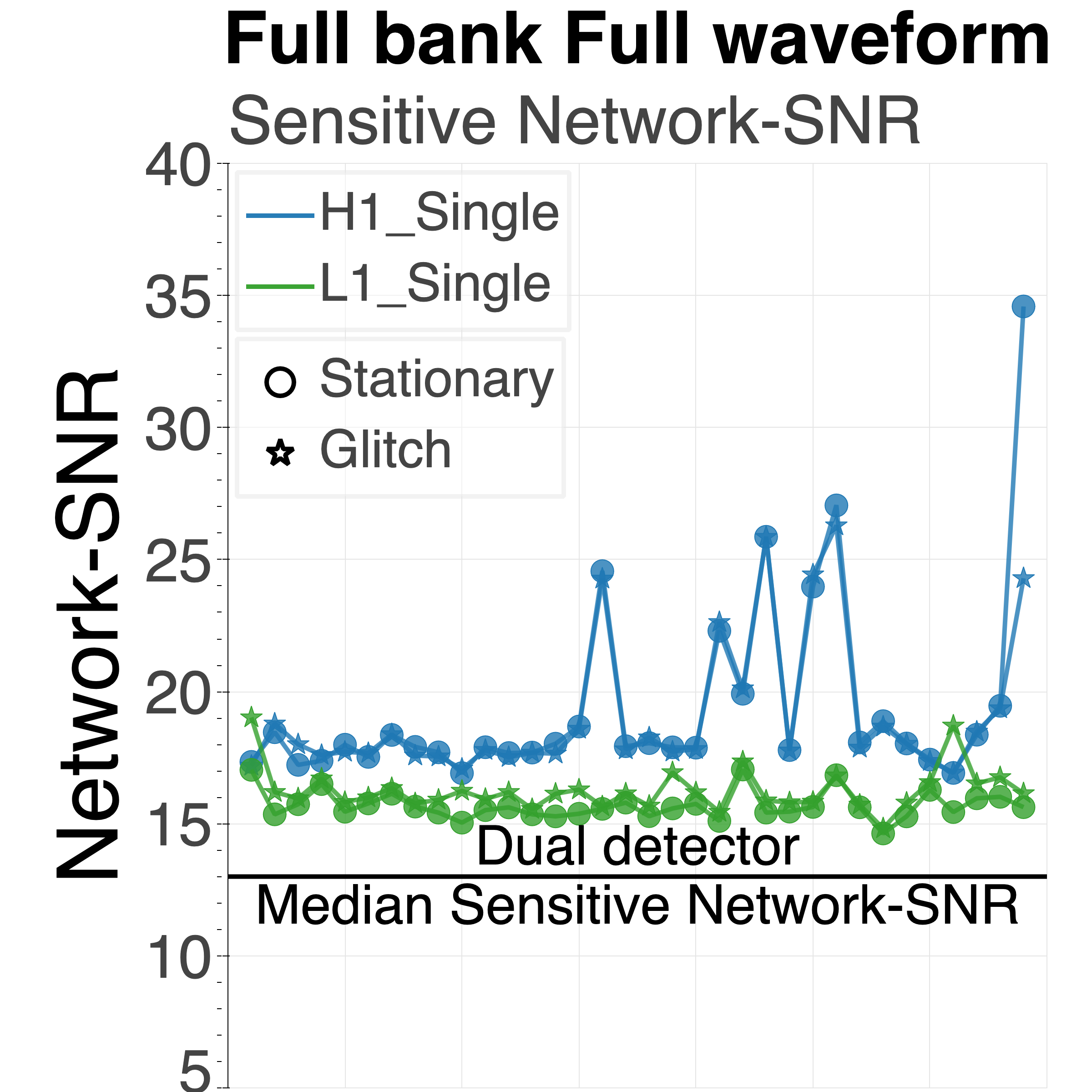}
    \end{subfigure}%
    \begin{subfigure}{0.25\textwidth}
        \includegraphics[width=\linewidth]{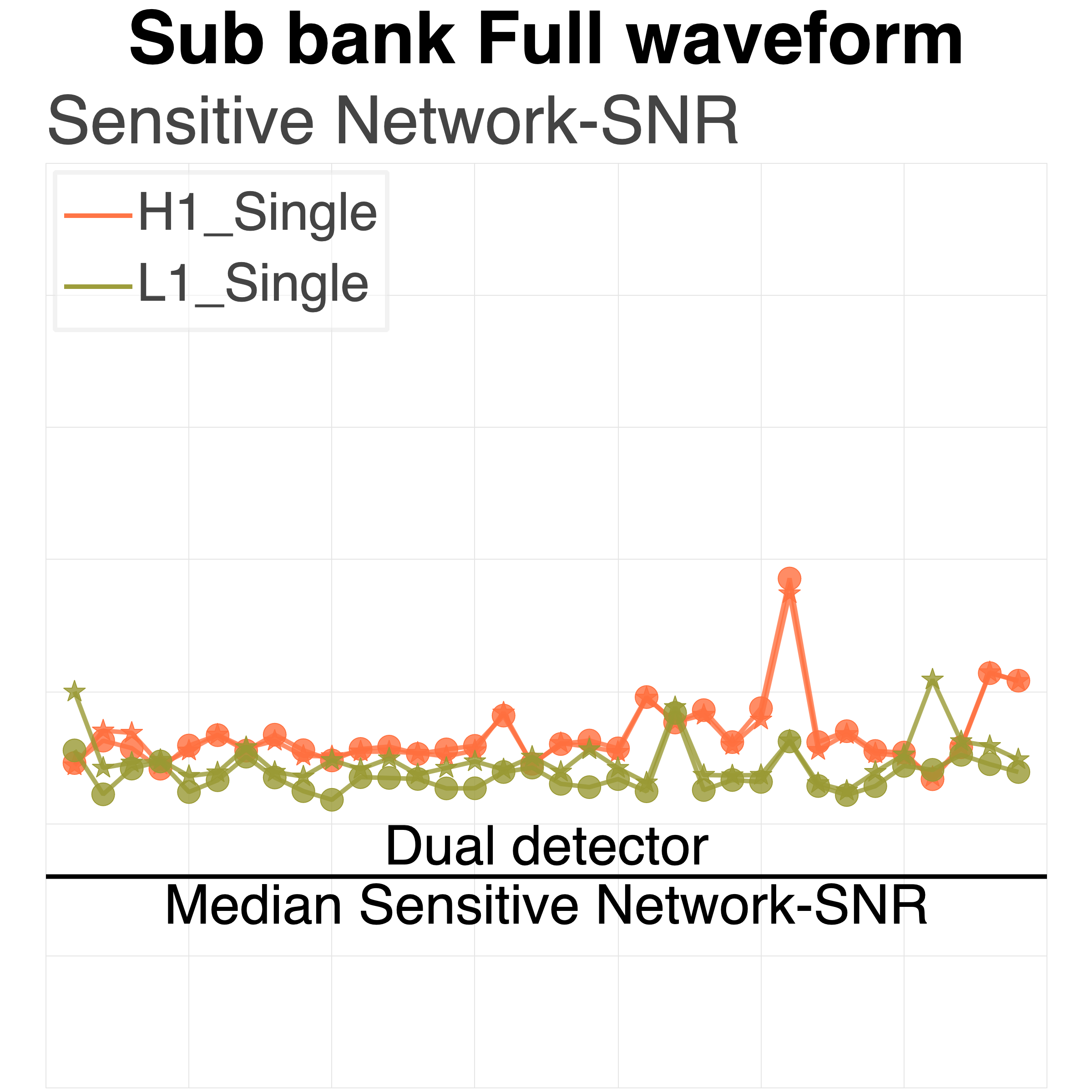}
    \end{subfigure}%
    \begin{subfigure}{0.25\textwidth}
        \includegraphics[width=\linewidth]{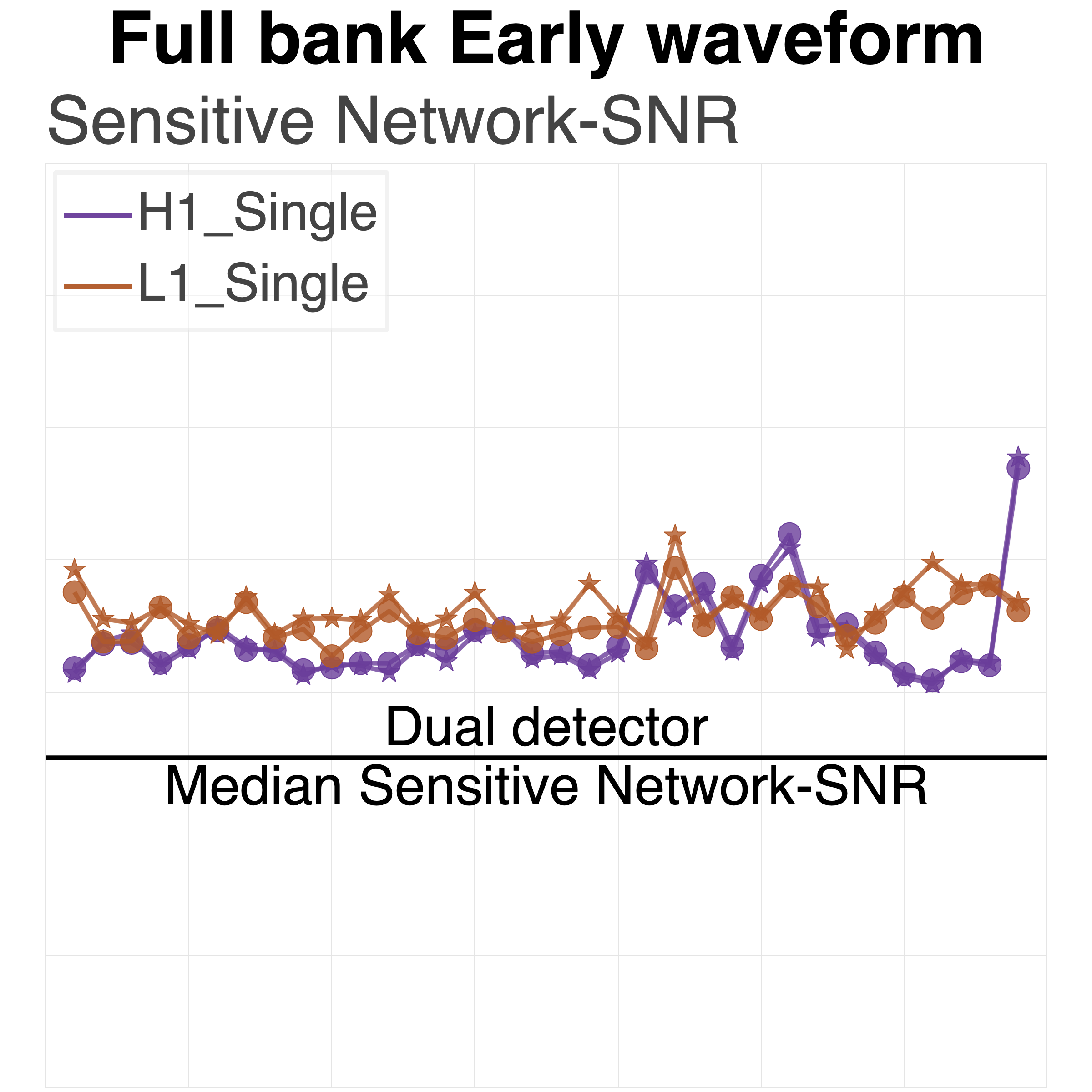}
    \end{subfigure}%
    \begin{subfigure}{0.25\textwidth}
        \includegraphics[width=\linewidth]{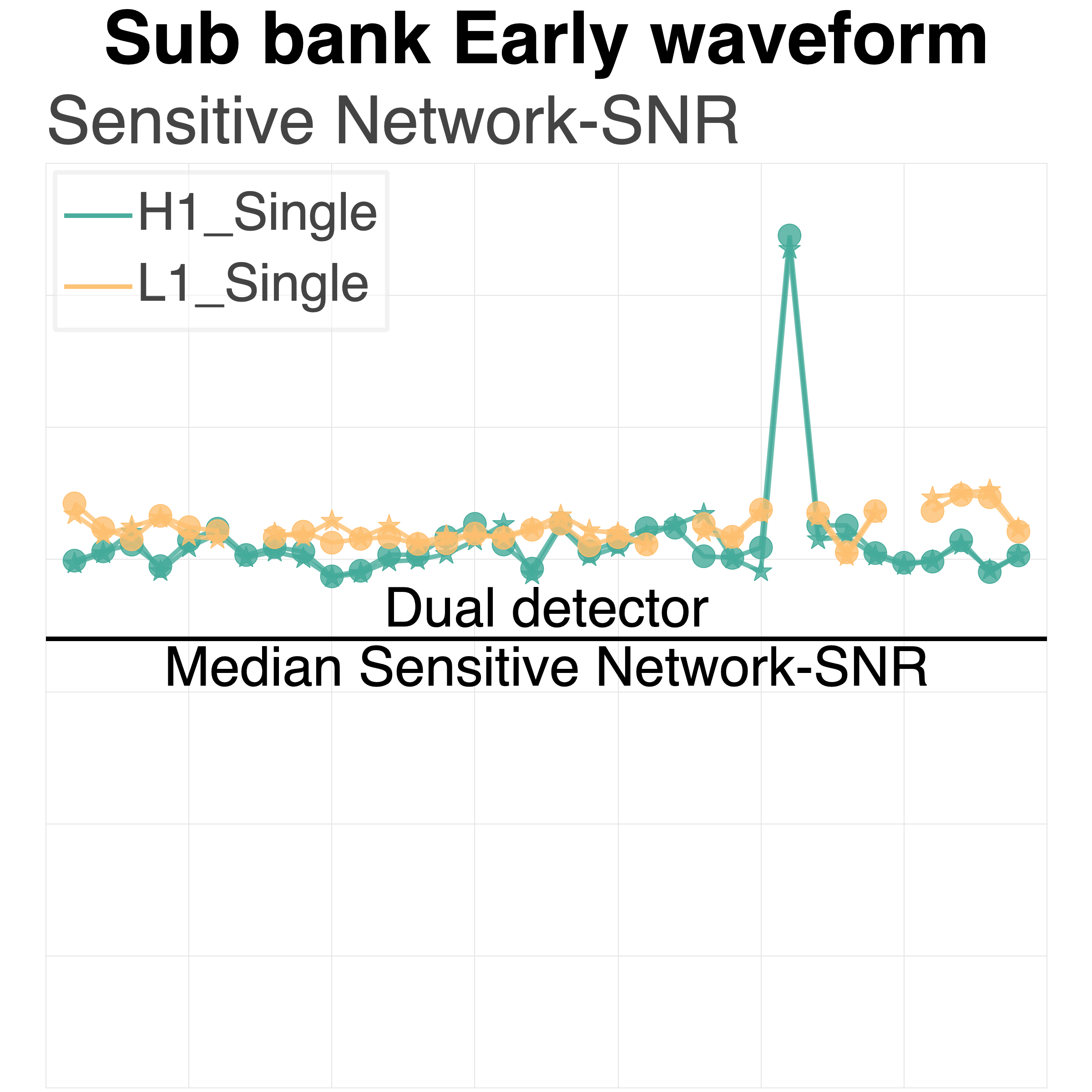}
    \end{subfigure}%

    \begin{subfigure}{0.25\textwidth}
        \centering
        \includegraphics[width=\linewidth]{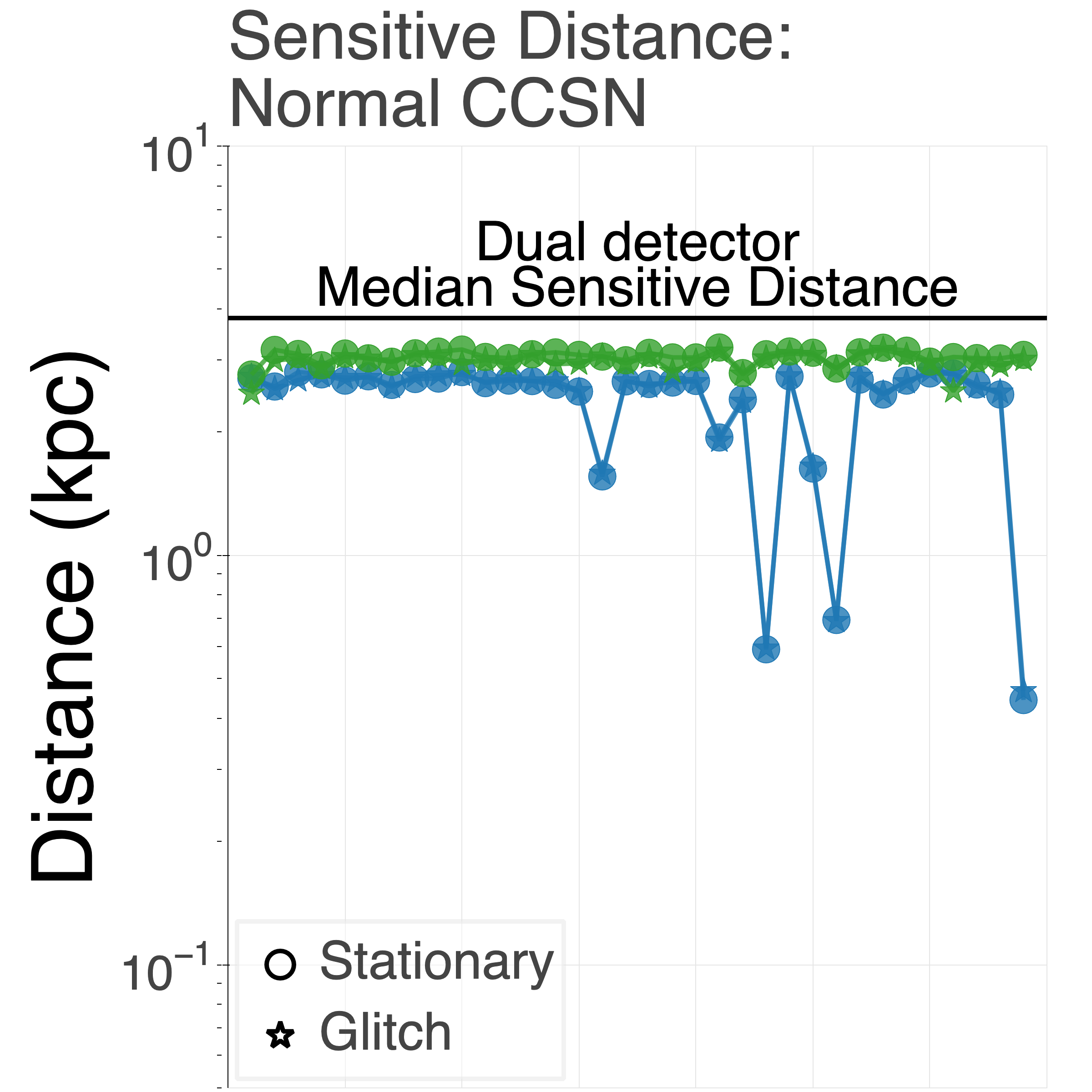}
    \end{subfigure}%
    \begin{subfigure}{0.25\textwidth}
        \includegraphics[width=\linewidth]{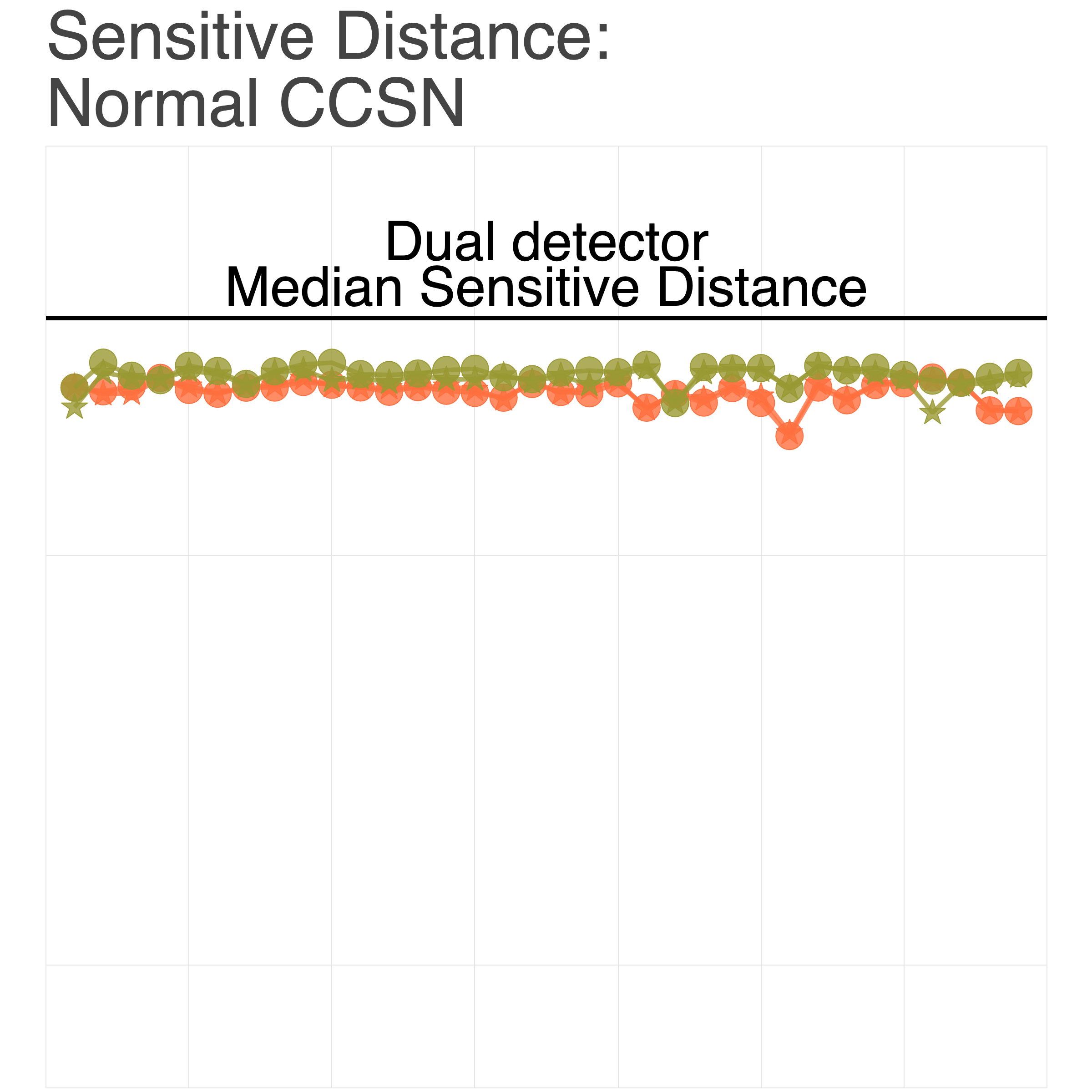}
    \end{subfigure}%
    \begin{subfigure}{0.25\textwidth}
        \includegraphics[width=\linewidth]{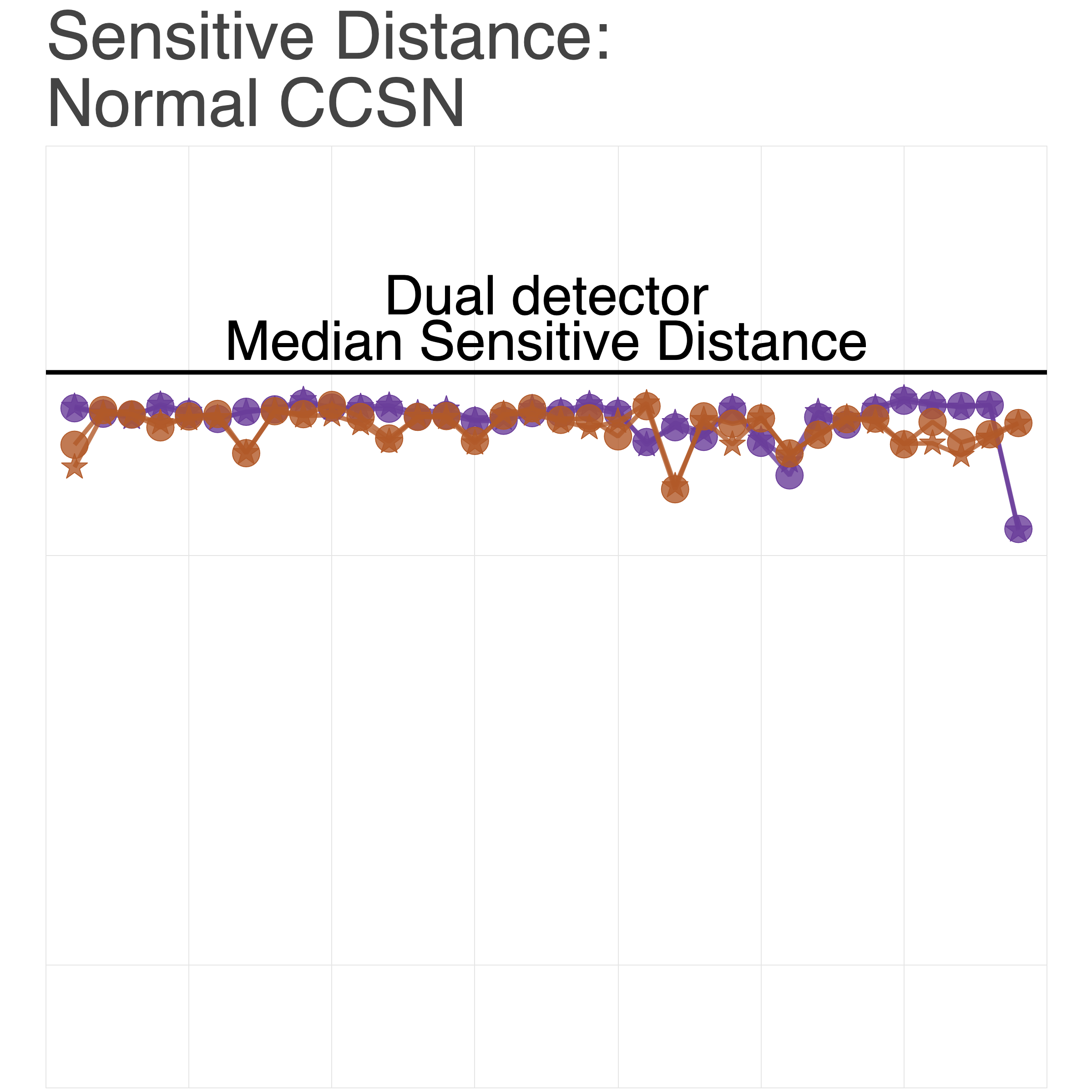}
    \end{subfigure}%
    \begin{subfigure}{0.25\textwidth}
        \centering
        \includegraphics[width=\linewidth]{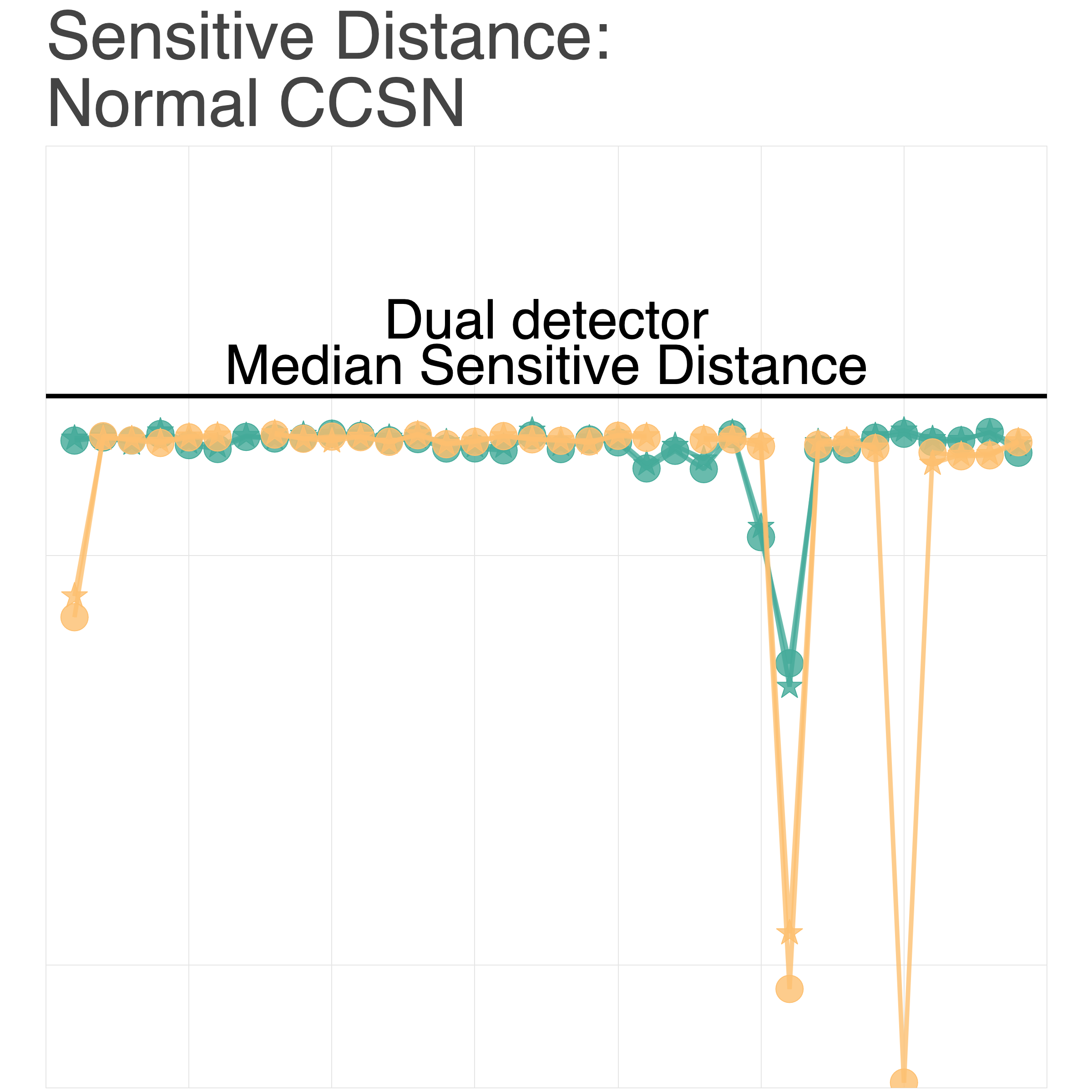}
    \end{subfigure}%
    
    \begin{subfigure}{0.25\textwidth}
        \centering
        \includegraphics[width=\linewidth]{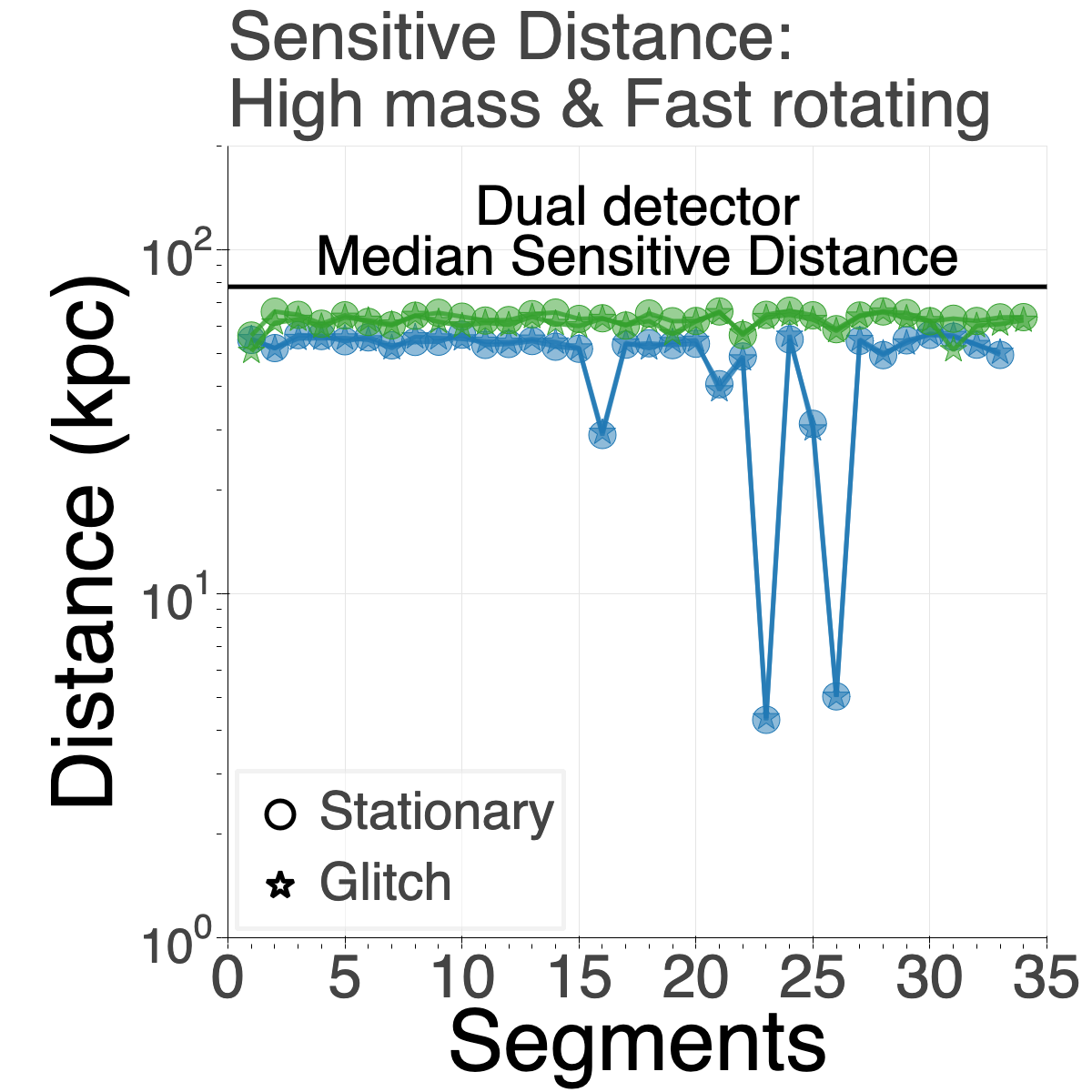}
    \end{subfigure}%
    \begin{subfigure}{0.25\textwidth}
        \includegraphics[width=\linewidth]{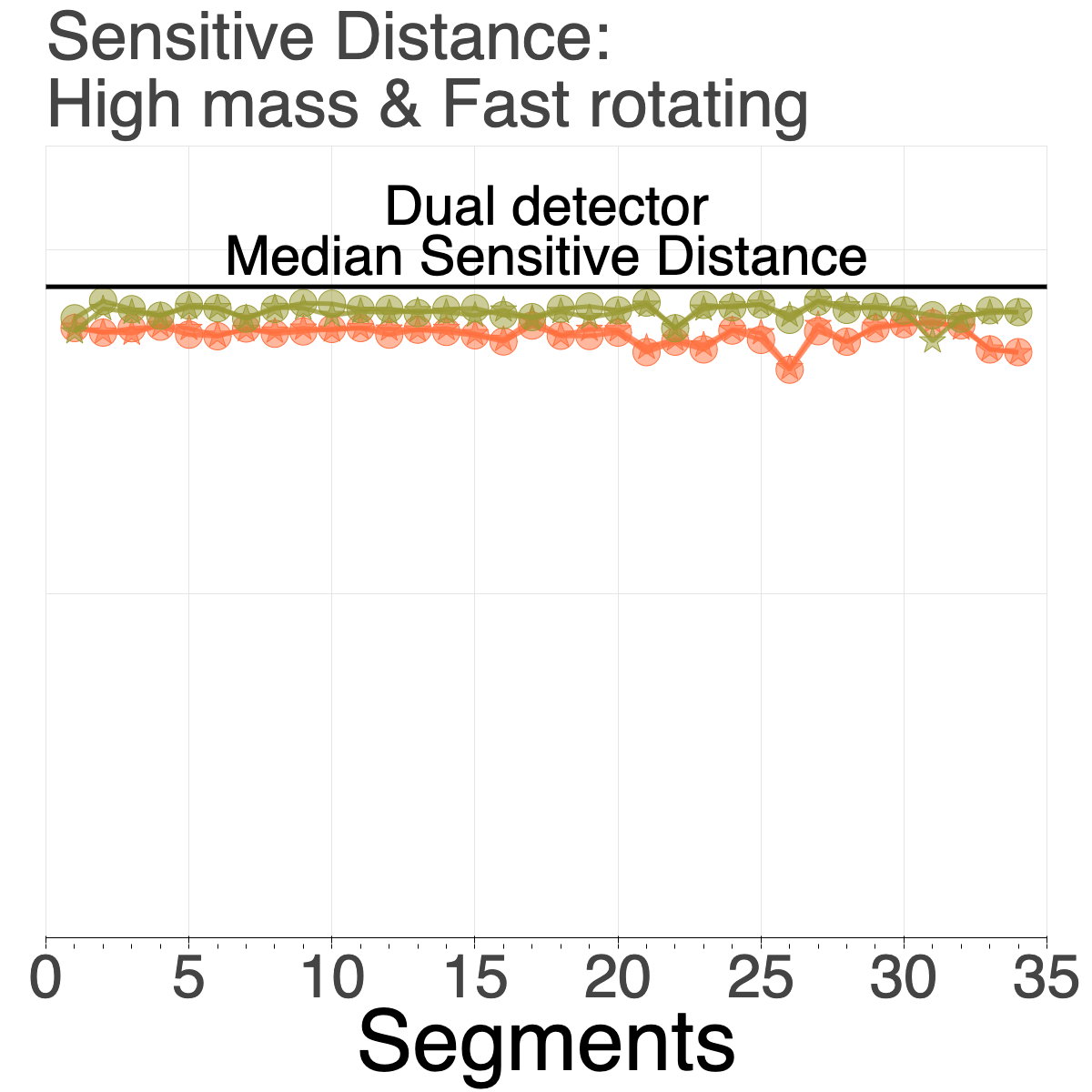}
    \end{subfigure}%
    \begin{subfigure}{0.25\textwidth}
        \includegraphics[width=\linewidth]{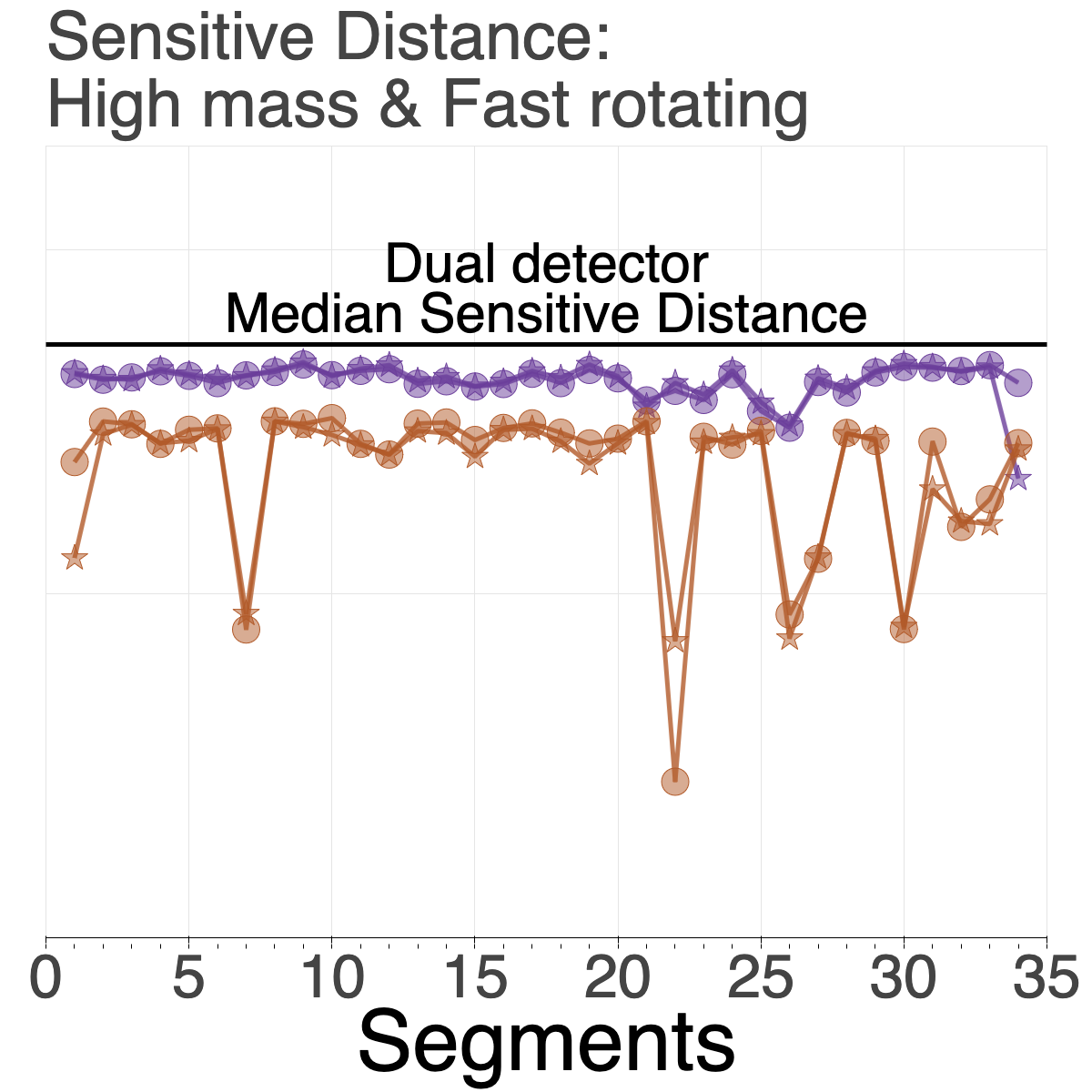}
    \end{subfigure}%
    \begin{subfigure}{0.25\textwidth}
        \centering
        \includegraphics[width=\linewidth]{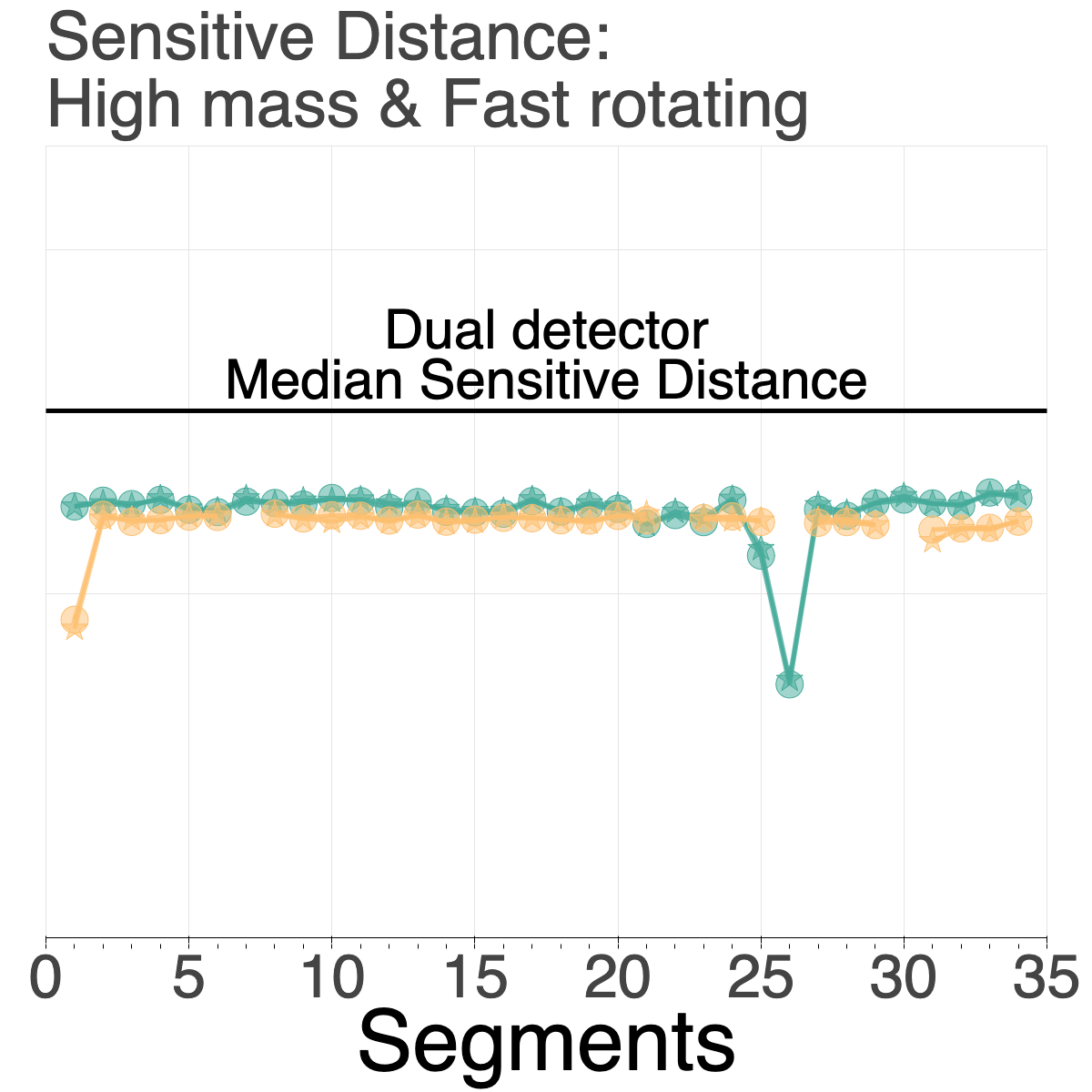}
    \end{subfigure}%
    \caption{
The top row shows the sensitive network SNR, while the middle and bottom rows display the sensitive distance for normal CCSNe and high-mass fast-rotating CCSNe, respectively. The colored lines indicate the sensitivity performance of H1-only or L1-only configurations under stationary and glitch-contaminated noise. The solid black horizontal line marks the dual-detector sensitivity baseline from the corresponding configuration. 
}
    \label{fig:segs_performance_single}
\end{figure*}

\subsection{Single-detector configuration}

Our model identifies CCSN triggers by learning waveform morphology from simulated templates, providing an alternative to coherence-based search strategies. Searching CCSN based on the wavefrom morphologies enables detection even when only a single detector provides analysis-ready data, thereby extending the effective search duty cycle. To leverage this capability, we modify the model input to utilize strain data from a single detector channel, while maintaining the architecture and hyperparameters unchanged. Training and validation datasets are generated using the same procedure illustrated in Figure~\ref{fig: train_data_workflow}, but restricted to single-detector noise and injections. Evaluation is performed on the same test dataset as in the dual-detector case, with inputs limited to data from one detector. 

The results of the single-detector configuration are shown in Figure~\ref{fig:segs_performance_single}. Relative to the dual-detector setup, sensitivity decreases by approximately $10$–$30\%$, indicating a modest performance loss while preserving effective detection capability. Models trained on the full waveform achieve comparable sensitivity regardless of whether a full or subset template bank is employed. Although glitches in single-detector backgrounds contribute to further reductions in sensitivity, the model retains meaningful detection ranges, with sensitive distances exceeding $1$–$2$~kpc for many CCSN waveform classes. These findings demonstrate that CCSN searches remain feasible during periods of single-detector operation, thus enhancing observational coverage.

\section{Discussion \label{sec_disscussion}}

Our supervised models demonstrate the ability to search for CCSN signals while simultaneously discriminating glitch transients. The comparison between Model\_FF and Model\_SF indicates that the framework remains robust when evaluated on CCSN waveform families that are not included in the training set. Figure~\ref{fig:method-result} shows that the waveforms not included during training Model\_SF achieve sensitivity comparable to Model\_FF. A consistent trend is also observed in the longevity study, as Figure~\ref{fig:longevity_performance} shows that the averaged sensitivity over the test segments follows a similar pattern for both models. Together, these results indicate that the classifier is not simply memorizing a fixed template set, but is instead learning morphological features that are shared across CCSN waveform families. 

The CCSN and glitch overlap tests further show that the model can identify CCSN injections in the presence of nonstationary artifacts, with only a modest reduction in sensitivity relative to the stationary background. This also implies that the decision boundary is not driven solely by the approximation of the glitch signals but also retains sensitivity to CCSN morphology. This result shows the benefit of incorporating CCSN waveform morphology, which helps preserve detection capability when glitches corrupt the strain data. 

We additionally tested configurations trained with minimal CCSN time-domain information. The performance of Model\_FE and Model\_SE indicates that CCSN triggering remains feasible under glitch overlap, but the sensitivity is substantially reduced compared with Model\_FF and Model\_SF. Moreover, comparing the sensitivity between Model\_FE and Model\_SE tells us that our current setup is unable to capture the shared core-bounce morphology effectively. As in Figure~\ref{fig:method-result}, the sensitivity-SNR has a larger deviation between the two models, which results in a worse average sensitivity along all test segments, as shown in Figure~\ref{fig:longevity_performance}. A likely cause of this performance reduction is the mismatch between the training signal duration and the inference window. For data preparation simplicity, we did not shorten the input kernel for Model\_FE and Model\_SE; consequently, the one-second inference kernel is much longer than the early waveform used during training, which is approximately 100 milliseconds. This introduces a large amount of redundant context that can dilute the early CCSN core-bounce morphology and effectively reduce the portion of CCSN presented to the ML model. Overcoming this issue will require aligning the kernel size to better match the injected signal duration, which needs dedicated hyperparameter and architecture tuning. 

We also explored an architectural modification that replaces dual-detector inputs with single-detector strain, which removes coherence information and requires corresponding changes to both noise and injection datasets. This configuration enables CCSN searches during times when only one interferometer is operating in science mode, thereby potentially increasing the search duty cycle. Despite the absence of coherence information, the model retains a meaningful detection range, as shown in Figure~\ref{fig:segs_performance_single}. In addition, models trained on either the full or the subset template banks with full-waveform injections achieve similar sensitivity, suggesting that the ML model can capture common features that appear across different CCSN templates. Because the input structure and available discriminants differ significantly from the dual-detector case, future work will focus on single-detector–specific architecture design and hyperparameter optimization to further improve sensitivity. 

Beyond model comparison, we also include a longevity test, which shows that while the performance is stable across most segments, sensitivity can degrade sharply in certain periods. These problematic segments coincide with cases in which the 95\% background-distribution threshold, shown in the top-left panel of Figure~\ref{fig:longevity_performance}, deviates from the normal trend. This pattern suggests that the loss in sensitivity is driven by changes in the background distribution rather than by the properties of injection data. We therefore retrained using noise from the first problematic segments and evaluated on subsequent segments. The retrained models improve performance over most of the problematic intervals, Figure~\ref{fig:segs_performance_retrain} for all four models. This result supports an operational strategy in which background monitoring is used to identify when retraining is necessary. More broadly, this monitoring–retraining–validation workflow shows that \texttt{CCSNet} is not only a static classifier, but an operational infrastructure that can update models and recover performance as detector conditions evolve. 

Several ML models have been developed as post-processors in GW detection pipelines to improve event significance. In the \cite{2011.13733} analysis, a ResNet-based network is implemented for noise rejection and CCSN classification. At a FPR of 5\% on O3 data, their CCSN sensitivity reaches a SNR $\sim$ 15 or a FAP of $\sim$ 10\% at TPR 50\%. The Mulasec \cite{105.084054} infrastructure employs a CNN for both classification and waveform reconstruction. This CNN is designed to cover a broad CCSN parameter space, enabling it to generalize across diverse signal morphologies. Their results show a sensitivity reaching SNR ~ 30 at a FAP of 10\%, with waveform reconstruction up to 1000 Hz. In \cite{2111.07219}, three classifier variants are investigated, among which SVMR achieves the best performance due to its demonstrated stability and robustness. These models are applied to data from different interferometer configurations, including L1, H1, and the combined H1+L1 network, achieving noise reductions of approximately 100–1000 on cWB~\cite{0802.3232} triggers. Overall, the results demonstrate that ML-based post-processing can substantially improve detection significance through effective discrimination of noise events. 

Due to the wide parameter space of CCSN models, GW signals exhibit significant variability across different simulations. As a result, the sensitivity profile of any given algorithm can vary considerably depending on the signal morphology. In the cWB O3 on-source targeted search analysis \cite{2305.16146}, the pipeline performed a sensitivity test on a broad range of CCSN templates, including  several 3-D waveforms we tested on as well, such as Anderson\_16  s11, O'Connor Mesa 20, Pan 2021 sr, and Powell 2020 s18. Across these four templates, they perform best in Powell`s waveform with a sensitive distance near half an order of magnitude compared to the other three waveforms. While in our analysis, the sensitivity of Pan 2021 sr  performs much better compared with most of the other CCSN followed by Powell 2020 s18, while Anderson\_16  s11 and O'Connor Mesa 20 remain as the less sensitive CCSNe. Both CCSN search analysis shows that sensitivity varies notably across different waveform families, highlighting the challenge to optimize the search algorithm for the full CCSN parameter space. 

Despite the promising performance of ML-based post-processing methods, several limitations remain. Owing to the large diversity of CCSN waveforms and the incomplete coverage of existing numerical simulations, it is not possible to guarantee that the current model fully spans the entire CCSN parameter space within the sensitivity range of the LVK detectors. As a result, model performance may degrade when encountering signal morphologies that differ substantially from those represented in the training dataset. In addition, the non-stationary nature of detector noise and evolving detector configurations requires continuous monitoring and periodic retraining of the model to maintain stable sensitivity and robust classification performance over time. These limitations highlight the need for broader waveform coverage, improved domain generalization, and adaptive training strategies for future CCSN GW searches. 

\section{Conclusion \label{sec_conclusion}}

While unmodeled burst-search pipelines are designed to target a broad range of astrophysical transients, \texttt{CCSNet} is developed specifically to identify CCSN waveform features under challenging detector conditions. These conditions include CCSN signals overlapping with glitch transients, searches with only limited time-domain CCSN information, and operation in single-detector configurations where multi-detector coherence information is unavailable.

The results demonstrate that \texttt{CCSNet} can effectively distinguish CCSN signals from detector noise artifacts, including cases in which signals overlap with glitches. Through targeted retraining strategies, the model performance can be maintained over extended observing periods, indicating that the framework is capable of adapting to evolving detector conditions. Across multiple training configurations, the network also shows the ability to generalize to waveform families that are not included in the training set. This behavior suggests that the classifier learns shared morphological characteristics of CCSN signals rather than memorizing waveform-specific features. However, models trained using injections restricted to the near core-bounce region exhibit noticeable sensitivity degradation, indicating that additional optimization is required to better capture generic CCSN features associated with the core-bounce phase.

\texttt{CCSNet} supports both dual-detector and single-detector operating modes, extending the effective search duty cycle beyond periods of coincident detector operation. In terms of applications, \texttt{CCSNet} can serve as a follow-up classifier for triggers identified by burst searches. In this study, we sampled all glitch transients labeled by Omicron for training, a strategy that was not optimized for any specific search pipeline. In future work, the glitch sampling strategy can be tailored to target noise transients that populate the tails of a given pipeline’s background distribution, thereby improving performance in the most relevant regime. 

Several limitations should nevertheless be noted. Due to the large diversity of CCSN waveform morphologies and the incomplete coverage of current numerical simulations, it cannot be guaranteed that the performance on the testing set can fully represent the CCSN parameter space accessible within the LVK sensitivity range. Consequently, the model performance may vary substantially across different waveform families. Additionally, the non-stationary nature of detector noise necessitates continuous monitoring and periodic retraining to maintain stable sensitivity throughout observing runs. Future work will therefore focus on expanding the training dataset to include broader CCSN waveform families and detector background conditions, while encouraging the model to learn more generic and physically robust CCSN features. Overall, \texttt{CCSNet} demonstrates that our ML classifier remains effective under realistic detector conditions, including glitch contamination and single-detector operation, serving as an additional method for CCSN identification in gravitational-wave burst searches.

\section{Acknowledgment}
We appreciate the help from the members of A3D3/ML4GW and the developers of \texttt{ml4gw}, who inspired us to design the code base with much better efficiency, scalability, and flexibility. In addition to the software support, we are also grateful to the computational resources provided by the support of the LIGO computational clusters and the KAMALAN-R750 server. Additionally, K.C.P. especially appreciates the support from the National Science and Technology Council of Taiwan through grants 113-2112-M-007-031.
\bibliography{CCSNet}

\end{document}